\documentclass[modern]{aastex631}

\usepackage{afterpage}
\usepackage{hyperref}

\shorttitle{SN2SNR D$^6$}
\shortauthors{Ferrand et al.}

\begin{document}

\title{The double detonation of a double degenerate system, \\from Type~Ia supernova explosion to its supernova remnant}

\author[0000-0002-4231-8717]{Gilles Ferrand}
\newcommand{\ABBL}{Astrophysical Big Bang Laboratory (ABBL), 
RIKEN Cluster for Pioneering Research,}
\newcommand{\iTHEMS}{RIKEN Interdisciplinary Theoretical and Mathematical Sciences Program (iTHEMS), \\ 
Wak\={o}, Saitama, 351-0198 Japan}
\affiliation{\ABBL}
\affiliation{\iTHEMS}

\email{gilles.ferrand@riken.jp}

\author[0000-0002-8461-5517]{Ataru Tanikawa}
\affiliation{Department of Earth Science and Astronomy, College of Arts and Sciences, The University of Tokyo, Meguro, Tokyo, 153-8902 Japan}

\author[0000-0002-3222-9059]{Donald C. Warren}
\affiliation{\iTHEMS}

\author[0000-0002-7025-284X]{Shigehiro Nagataki}
\affiliation{\ABBL}
\affiliation{\iTHEMS}

\author[0000-0001-6189-7665]{Samar Safi-Harb}
\affiliation{Department of Physics \& Astronomy, University of Manitoba, Winnipeg, MB R3T 2N2, Canada}

\author[0000-0002-1796-758X]{Anne Decourchelle}
\affiliation{Université Paris-Saclay, CEA, CNRS, AIM, F-91191 Gif-sur-Yvette, France}
\affiliation{Université de Paris, AIM, F-91191 Gif-sur-Yvette, France}

\accepted{by ApJ}

\begin{abstract}
Type~Ia supernovae (SNe) are believed to be caused by the thermonuclear explosion of a white dwarf (WD), but the nature of the progenitor system(s) is still unclear. Recent theoretical and observational developments have led to renewed interest in double degenerate models, in particular the ``helium-ignited violent merger'' or ``dynamically-driven double-degenerate double-detonation'' (D$^6$). 
In this paper we take the output of an existing D$^6$ SN model and carry it into the supernova remnant (SNR) phase up to 4000 years after the explosion, past the time when all the ejecta have been shocked. Assuming a uniform ambient medium, we reveal specific signatures of the explosion mechanism and spatial variations intrinsic to the ejecta. 
The first detonation produces an ejecta tail visible at early times, while 
the second detonation leaves a central density peak in the ejecta that is visible at late times.
The SNR shell is off-centre at all times, because of an initial velocity shift due to binary motion.
The companion WD produces a large conical shadow in the ejecta, visible in projection as a dark patch surrounded by a bright ring. 
This is a clear and long-lasting feature that is localized, and its impact on the observed morphology is dependent on the viewing angle of the SNR.
These results offer a new way to diagnose the explosion mechanism and progenitor system using observations of a Type~Ia SNR.
\end{abstract}

\keywords{supernovae, supernova remnants, hydrodynamical simulations}

\section{Introduction} 
\label{sec:intro}

Supernovae (SNe) mark the end of the life of stars, and are one key step in the life cycle of elements in the Galaxy. Type Ia SNe have attracted a lot of attention, for enabling the discovery of cosmic acceleration \citep[see][for a review]{Garnavich2017DiscoveryAcceleration}. SNe Ia are thought to be the thermonuclear explosion of a white dwarf (WD), although they come in many types \citep{Jha2019Observationalsupernovae} and their progenitor system(s) are still unclear \citep{Ruiter2020TypeOrigin}. Many scenarios have been proposed for the explosion \citep{Hillebrandt2013TowardsObservations}, a basic idea being that the WD is part of a binary system so that it can increase its mass via accretion or merger. 

There is no consensus on the nature of companion stars of exploding WDs. The companion star may be a non-degenerate star -- the single degenerate scenario \citep{Whelan1973BinariesI, Nomoto1982AccretingSupernovae}, another WD -- the double degenerate scenario \citep{Iben1984Supernovaemass, Webbink1984Doublesupernovae}, or the core of an asymptotic giant star -- the core degenerate scenario \citep{Kashi2011Acircumbinarysupernovae}.
Recent observations have put some constraints on the single degenerate scenario. 
No red-giant star has been found in the pre-explosion images of SN 2011fe and 2014J \citep[][respectively]{Li2011NearbySample, Kelly2014NearbySample}, which are the nearest type Ia SNe observed in the past decade. 
No surviving stellar companions were convincingly detected in the remnants of nearby Ia SNe \citep[see][for a review]{Ruiz-Lapuente2019SurvivingObservations}.
On the other hand, some type Ia SNe may indicate the presence of non-degenerate companion stars, such as PTF11kx \citep{Dilday2012PTF11kx:Progenitor}, iPTF14atg \citep{Cao2015StrongSupernova}, and SN 2012cg \citep{Marion2016InteractionWithBinaryCompanion}. These observational results suggest that type Ia SNe may have multiple origins \citep{Hillebrandt2013TowardsObservations}. 

The double degenerate scenario is one of the promising scenarios for type Ia SNe. In this scenario, the exploding WD may be a near-Chandrasekhar mass explosion with high central density, or a sub-Chandrasekhar mass explosion with low central density. Near-Chandrasekhar mass explosion has been thought to be difficult theoretically, because such a WD converts its composition from carbon-oxygen to oxygen-neon-magnesium \citep{Saio1985OxygenNeonMagnesium, Schwab2012MagneticViscosity, Ji2013MagneticViscosity}, and collapses to a neutron star rather than exploding as a type Ia SN \citep{Nomoto1991AccretionInducedCollapse}. Various sub-Chandrasekhar mass explosion models have been suggested. Although the violent merger model can successfully produce type Ia SNe \citep{Pakmor2011ViolentSupernovae, Pakmor2012NormalBinaries, Tanikawa2015ViolentMerger}, the total mass of two WDs should be larger than the Chandrasekhar mass \citep{Sato2015ViolentMerger, Sato2016ViolentMerger}, which means that the event rate is smaller than the type Ia SN rate \citep{Maoz2014ObservationalSupernovae}. Another model, called the ``helium-ignited violent merger'' or ``dynamically-driven double-degenerate double-detonation'' (hereafter, D$^6$) allows the total mass of the two WDs to be less than the Chandrasekhar mass \citep{Guillochon2010SurfaceDetonations, Pakmor2013Helium-ignitedSupernovae}. The D$^6$ scenario is supported by the recent discovery of a hyper-velocity WD by \cite{Shen2018ThreeSupernovae}, which is expected to be left as a by-product of the explosion. Thus, the D$^6$ model has become a promising model for type Ia SNe.

This has motivated many researchers to examine the double detonation of a sub-Chandrasekhar-mass WD with respect to its explodability, nucleosynthesis, and observability \citep[e.g.][]{Shen2018Sub-Chandrasekhar-massRevisited, Shen2021RadiativeTransfer, Gronow2020SNeMechanism, Gronow2021DoubleDetonations, Polin2019ObservationalPredictions, Polin2021NebularModels,Pakmor2021HybridHeCO}. These studies have mainly focused on the exploding WD, not the surviving WD. \cite{Papish2015TheSupernova} and \cite{Tanikawa2018Three-DimensionalCompanion, Tanikawa2019Double-DetonationMaterials} have shown that D$^6$ SN ejecta exhibit non-spherical features because of the presence of the surviving WD. If such non-spherical features can survive until the remnant phase, it would be inconsistent with spherically symmetric objects. On the other hand, SN ejecta may become spherical through interaction with the interstellar medium (ISM). In addition, the chemical compositions of D$^6$ ejecta would be different from those of SN Ia ejecta with a non-degenerate companion star. In the D$^6$ scenario, ejecta can contain a significant amount of carbon and oxygen, stripped from the companion WD \citep{Tanikawa2018Three-DimensionalCompanion, Tanikawa2019Double-DetonationMaterials}. If the companion star is non-degenerate, the ejecta may contain hydrogen stripped from a main-sequence or red-giant star, or helium in the case the companion is a helium star \citep{Marietta2000TypeIAConsequences, Pan2012ImpactScenario, Liu2013TheScenario, Boehner2017ImprintsCompanions}. It is unknown whether this difference is observable or not after SN ejecta interact with the ISM.

The supernova remnant (SNR) is the phase following the SN, made from the interaction of the ejecta with the circumstellar and/or interstellar matter \citep{Reynolds2017DynamicalRemnants}. In a young, ejecta-dominated SNR, the shell of shocked matter is bounded by two shocks: the forward shock (FS) developing ahead of the supersonic ejecta, and the reverse shock (RS) forming inside the ejecta as they are decelerated. The shocked matter (shocked ejecta and shocked ISM) is heated to X-ray emitting temperatures, for thousands of years \citep{Vink2017X-RayRemnants}. The interface between the ejecta and the ISM, a contact discontinuity (CD), is unstable and subject to the Rayleigh-Taylor instability (RTI). This shapes the SNR independently of its initial conditions (the explosion) and boundary conditions (the ISM). 

In this work we are specifically investigating the impact of the initial explosion on the SNR morphology. A~D$^6$ SN is rather symmetric, except for the marked imprint from the companion WD. 
In the study of thermonuclear explosions in binary systems, there has been a number of works that considered the interaction of the ejecta with the companion star, in terms of the impact on the SN emission, or the fate of the companion \citep[e.g.][]{Marietta2000TypeIAConsequences,Pakmor2008TheimpactCompanions,Kasen2010SeeingStar,Liu2013TheScenario,Maeda2014SignaturesSupernovae,Boehner2017ImprintsCompanions,Dessart2020SpectralSupernova,Zeng2020TheStar,Liu2021Long-termSupernovae}. 
There have also been several searches for surviving companions in SNRs \citep{Ruiz-Lapuente2019SurvivingObservations}, but looking for normal stars -- an exception being \cite{Kerzendorf2018A1006}, who looked for a WD too, motivated by the theoretical study of \cite{Shen2017WaitDecays}. A~feature of the D$^6$ scenario is that the surviving companion is itself a WD.\footnote{A~companion WD could also exist in a single degenerate scenario with the spin-up/spin-down mechanism \citep{Justham2011Single-degenerateContamination, DiStefano2011SpinUpSpinDown, Hachisu2012SpinUpSpinDown, Benvenuto2015SpinUpSpinDown}.}
In contrast only a~few works have looked into the signature of a companion on the SNR --~and all assuming normal stars for the companion \citep{Lu2011TheCompanion,Vigh2011AsymmetriesRemnants,Garcia-Senz2012IsRemnants,Gray2016ShadowsRemnants}. In this paper we investigate for the first time the imprints on the SNR with a WD companion, and follow the evolution up to when all the ejecta have been shocked.

Our approach is to run numerical simulations from the 3D SN to the 3D SNR. We already applied it to the canonical model for a Type Ia SN: the explosion of a Chandrasekhar-mass WD \cite[][hereafter Paper~I]{Ferrand2019FromExplosion}, examining the impact of the ignition and of the propagation of the flame \cite[][hereafter Paper~II]{Ferrand2021FromModels}. In Paper~I we showed that, assuming a uniform ambient medium, the impact of the SN on the SNR may still be visible after hundreds of years. Furthermore, in Paper~II we showed that the details of the explosion mechanism matter: the more asymmetric ignition setups produce more asymmetric remnants, and in a way that is different for deflagrations versus detonations. In this paper we are investigating a different kind of Type Ia model: the double detonation of a WD in a double degenerate system.

The paper is organized as follows. In Section~\ref{sec:method} we summarize the methods for carrying out the numerical simulations and analyzing their results. In Section~\ref{sec:results} we present the results on the SNR morphology, using a selection of representative maps. In Section~\ref{sec:discussion} we discuss implications for observations of young, nearby SNRs. In Section~\ref{sec:conclusion} we conclude and outline our perspectives with the $D^6$ model. 
\\

\section{Method} 
\label{sec:method}

\afterpage{

\begin{table}[t!]
\vspace{5mm}
\begin{center}
\begin{tabular}{ r|l } 
primary's mass & $M_1 = 1.0$~$M_\odot$ (95\% CO + 5\% He) \\ 
secondary's mass & $M_2 = 0.6$~$M_\odot$ (100\% CO) \\ 
orbital separation & $d_{12} = 1.6 \times 10^4$ km \\
primary's orbital velocity & $v_1 = 1\,100$~km s$^{-1}$ \\ 
secondary's orbital velocity & $v_2 = 1\,800$~km s$^{-1}$ \\ 
angular size of secondary seen from primary & $2\theta_2 = 44^{\circ}$, $\Omega_2 = 0.46$~sr \\
\hline
ejecta mass & $M_\mathrm{ej} = 0.97$~$M_\odot$ \\
total kinetic energy & $E_\mathrm{SN} = 1.11\times10^{51}$~erg \\
ISM density & $\rho_\mathrm{ISM} = 0.1$~m$_p$.cm$^{-3}$ \\
\end{tabular}
\end{center}
\caption{Model parameters
\label{tab:params}}
\vspace{5mm}
\end{table}

\begin{figure}[h!]
\centering
\includegraphics[width=0.9\textwidth]{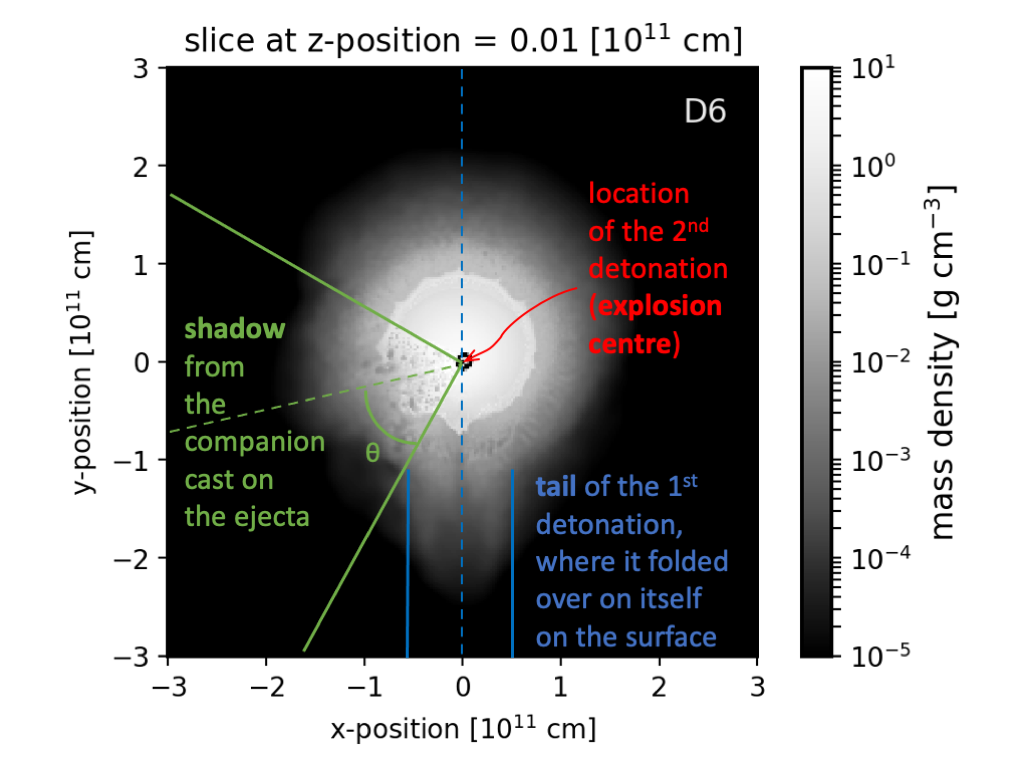}
\caption{Slice of the ejecta of the D$^6$ explosion, annotated with its specific features. This figure was made from the output data at 50~s of the SN simulation published in \cite{Tanikawa2018Three-DimensionalCompanion}, that are the input data of the SNR simulation presented in this paper. The quantity plotted is the mass density of the ejecta of the exploded primary WD. The slice is taken in the x-y plane, through the origin of the z-axis. The tail visible at the bottom (blue outline) is where the first detonation folded over on itself, on the surface of the primary WD, opposite to the ignition point. The second detonation, that destroyed the primary WD, happened near its centre (red arrow). The shadow that was cast in the ejecta of the primary WD by the presence of the secondary WD is visible on the bottom left side (green outline). It has a conical shape, of half-opening angle $\theta$. The inner part of it, along the cone axis, is contaminated with material stripped from the secondary WD. (At the scale of this plot, the two WDs can barely be distinguished, their separation is of the order of the thickness of the solid lines.)
\label{fig:sketch}}
\end{figure}

\clearpage
}

Our simulation is done in two steps, the SN explosion per se, already published, and the subsequent SNR evolution, presented in this paper.

\paragraph{The supernova}
Our D$^6$ SN model is described in detail in \cite{Tanikawa2018Three-DimensionalCompanion}. The main model parameters are summarized in Table~\ref{tab:params}. The simulation was made with a smoothed-particle hydrodynamics (SPH) code, using a Helmholtz equation of state \citep{Timmes2000HelmholtzEoS}, and coupled with the nuclear reaction network Approx13 \citep{Timmes2000Approx13}. As for the initial conditions, we considered a binary system with WDs of mass $1.0 M_\odot$ for the primary and $0.6 M_\odot$ for the secondary. According to \cite{Shen2018Sub-Chandrasekhar-massRevisited} (see their Figure~5) a~$1.0 M_\odot$ WD produces about the mass of $^{56}$Ni required to power a typical SN~Ia, while lower masses would produce under-luminous events and higher masses would produce over-luminous events. As for the mass of the secondary, the way it impacts the explodability of the primary is not settled, but it should not affect the result from the explosion of the primary. 
The primary is made of 95\% CO in its core and 5\% He in its shell, while the secondary is pure CO with equal C and~O, in mass (there is the possibility that the secondary also has an He layer). Although the exact composition of the primary WD does affect the yields somewhat, it does not alter the total kinetic energy of the ejecta \citep{Sim2010DetonationsDwarfs}, which controls the dynamics of the remnant phase.
The two WDs orbit each other with a semi-major axis $1.6 \times 10^4$ km. The first detonation is ignited by adding a hot spot on the surface of the primary. The hot spot is located in the orbital plane of the binary system, and in the propagating direction of the primary WD.\footnote{\cite{Guillochon2010SurfaceDetonations} obtained that the first detonation starts at the endpoint of the accretion stream, but \cite{Pakmor2013Helium-ignitedSupernovae} obtained that it starts far from this endpoint, which suggests that it could start at any point on the orbital plane.} The first detonation develops over the He layer, and folds over on itself in t = 1.25~s, making a splash at the opposite point. The shock is then channeled into the WD, it converges at the centre at t = 1.625 s, triggering the second detonation, the thermonuclear explosion that disrupts the primary WD. The explosion generates a kinetic energy of $\simeq 10^{51}$~erg, and synthesizes a mass of $^{56}$Ni of 0.54 $M_\odot$. 

The interaction of the primary WD's ejecta with the secondary WD produces distinctive features in otherwise spherically symmetric ejecta, as seen in Figure~5 in \cite{Tanikawa2018Three-DimensionalCompanion}, and our annotated Figure~\ref{fig:sketch}. The main feature is a wide shadow in the ejecta (solid angle $\Omega = 1.8$~sr, of conical shape with half opening angle $\theta=44.5^\circ$), a~region of lower density, more irregular material, that was cast by the companion. An interactive 3D model of the ejecta is available online,\footnote{on Sketchfab at \href{https://skfb.ly/6VXwU}{skfb.ly/6VXwU}} made from iso-contours of the mass density, that shows the conical shadow. A~second feature is a narrow stream ($\Omega = 0.21$~sr, $\theta=14.8^\circ$) of material that was stripped from the companion, in particular low-velocity carbon and oxygen moving at $\simeq 3000$ km.s$^{-1}$. Such low-velocity oxygen can be seen in the nebular phase of SN 2010lp \citep{Taubenberger2013SN2010lp, Kromer2013SN2010lp} and iPTF14atg \citep{Kromer2016iPTF14atg}, two SN 2002es-like sub-luminous type Ia SNe.
Also visible in Figure~\ref{fig:sketch} in the outer, less dense region, is an extended tail, along the line where the first detonation folded over on itself on the surface of the primary WD. Another feature of the D$^6$ explosion is that the ejecta have a velocity shift of $\sim 1000$ km.s$^{-1}$. In our simulation the secondary WD survives the explosion of the primary (it may or may not explode, see \citealt{Tanikawa2019Double-DetonationMaterials}). It has a velocity $\simeq 1\,700$ km s$^{-1}$, that is travelling 1~pc every 575~yr, in a direction perpendicular to the axis of the conical shadow, and it has a peculiar composition from ejecta contamination.

\paragraph{The supernova remnant}
Results of the SN simulation at 50~s were mapped to a Cartesian grid of physical size $6\times10^{11}$~cm with spatial resolution $256^3$, and loaded into our custom version of the Eulerian hydrodynamics code Ramses \citep{Teyssier2002CosmologicalRAMSES} to conduct the SNR simulations. The ability of the code to follow SNR dynamics was demonstrated in our earlier works focused on particle acceleration \citep{Ferrand20103DAcceleration,Ferrand2012Three-dimensionalAcceleration,Ferrand2014Three-dimensionalAcceleration}. For $D^6$, the companion itself is not included in the SNR simulation (its size is $\lesssim 2\times10^9$ cm, about the size of a single grid cell), only its impact on the ejecta is considered (the direct interaction between the companion WD and the ejecta is finished by that time). The method is the same as in Papers~I and~II. Considering the initial quasi self-similarity, we first scale up the hydro profiles from 50~s to 1~day, and start the SNR simulation at 1~day. We follow the evolution up to 4000 years after the explosion, which is after the RS has shocked all of the ejecta and reached the SNR centre. This is a more advanced age than in the previous papers, because we observed longer-lasting features with~D$^6$. The SNR remains in an adiabatic phase of evolution, with our parameters the radiative snowplow phase would start at about $50\,000$ yr \citep{Cioffi1988DynamicsRemnants}.
The SNR simulation is performed in a comoving grid: factoring out the global expansion of the SNR over that period of time allows us to focus numerical resolution on the dynamics of the shocked ejecta.\footnote{Numerical resolution affects our ability to capture the RTI: the instability always grows from the smallest scales available, up to larger scales over time. A~resolution study was performed for the N100 model, and the overall trends on maps and angular spectra were found to be robust -- see the discussion in Section~4.1 in Paper~I.} For reference the box size is $4.5\times10^{-4}$ pc at 1~day, 5.6~pc at 100~yr, 40~pc at $4\,000$~yr. In order to separate the effects of the explosion and of any structure in the ambient medium, we assume a homogeneous ISM, of number density 0.1~cm$^{-3}$ to have dynamics roughly similar to Tycho's SNR. The overall hydrodynamic evolution of the SNR is controlled by three characteristic scales for radius, velocity, and time \citep{Dwarkadas1998InteractionSurroundings,Warren2013Three-dimensionalSNRs}:
\begin{eqnarray}
r_\mathrm{ch} &=& \left(\frac{3 M_\mathrm{ej}}{4 \pi \rho_\mathrm{ISM}}\right)^{1/3} \simeq 4.54\mathrm{~pc} , \label{eq:r_ch} \\
u_\mathrm{ch} &=& \left(\frac{2E_\mathrm{SN}}{M_\mathrm{ej}}\right)^{1/2} \simeq 10\,700\mathrm{~km.s}^{-1}, \label{eq:u_ch} \\
t_\mathrm{ch} &=& \frac{r_\mathrm{ch}}{u_\mathrm{ch}} \simeq 412\mathrm{~yr}, \label{eq:t_ch}
\end{eqnarray}
where $M_\mathrm{ej}$ is the mass of the ejecta, $E_\mathrm{SN}$ the (kinetic) explosion energy, and $\rho_\mathrm{ISM}$ the mass density of the ISM.

The analysis of the SNR morphology is performed as explained in the previous papers. At each step we track three surfaces of particular interest: RS, CD, FS. At runtime we extract in 3D the surface of each wave front, and record its radius from the explosion centre. Treating the latter as a function on a sphere, we expand its relative variations in spherical harmonics and compute the angular power spectrum. The simulation was also done with smooth initial conditions (labelled 1Di, vs 3Di for the full initial conditions), made by averaging the mass density over all angles, in order to assess what happens from the SNR phase only. 
\\

\section{Results} 
\label{sec:results}

\begin{figure}[t!]
\centering
\includegraphics[clip,trim={0 60 160 0},width=0.8\textwidth]{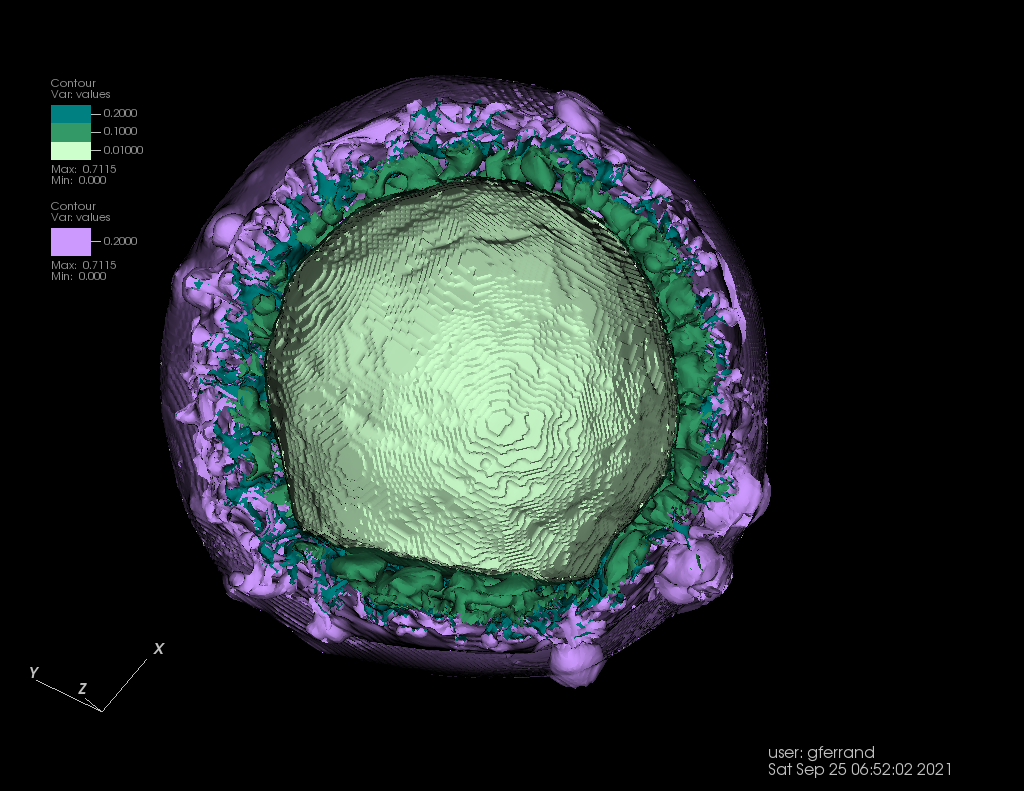}\\
\includegraphics[clip,trim={0 60 160 0},width=0.8\textwidth]{figure2A.png}
\caption{3D view of the D$^6$ SNR at 500~yr. The solid contours are iso-contours of the mass density. Green colours in the inner layers correspond to shocked ejecta, while purple colours in the outer layers correspond to shocked ambient matter. In both panels the contours are clipped to one half of the SNR to see its interior.
In the bottom panel, a volume rendering of the mass density of the shocked ejecta is included, over the entire surface of the SNR. 
\label{fig:3Dplot}}
\end{figure}

Our simulation shows the expected dynamical evolution for a young SNR: the formation of a shell of dense and hot matter bounded by the RS and the FS. This is the region where thermal emission is produced in the X-ray band, that can be observed with space-borne telescopes. In Figure~\ref{fig:3Dplot} we show a 3D view of the SNR shell, at an age of 500~yr for illustration. On this plot only the shocked matter is shown, the ejecta shocked by the RS (green hues) and the ISM shocked by the FS (purple hue). We~use both iso-contours and a volume rendering of the mass density. Part of the shell has been removed to show its interior. The typical field of Rayleigh-Taylor fingers is visible all around. The shell is globally symmetric except for a flat edge of the RS (at the bottom of the plot), with reduced RTI between the shocks. As we will see more clearly in the subsequent figures, this is the imprint from the shadow cast by the companion WD in the ejecta of the exploded WD. We also note a region with more marked fingers (toward the bottom right corner of the plot), this corresponds to the tail where the first detonation folded over on itself. 
Another 3D model of the SNR is available online,\footnote{on Sketchfab at \href{https://skfb.ly/oqoGX}{skfb.ly/oqoGX}} made from iso-contours of the ejecta tracer, that shows the morphology of the RT fingers in the shell of shocked matter, and is interactive. 

In the following, in order to explain the morphological evolution we use three different kinds of plots: i) 2D slices of the mass density, of all the matter, to reveal the inner structure of the SNR (Figure~\ref{fig:map_cut_rho}); ii) 2D projections along a line of sight of the density squared, that serves as a proxy of the thermal X-ray emission (Figure~\ref{fig:map_prj_f}); and iii) spherical maps and angular spectra, that show the surface of the wave fronts and quantify their angular content (Figures~\ref{fig:healpix_CD}, \ref{fig:healpix_RS}, \ref{fig:healpix_FS}). To make it easier to identify the features from the explosion, we generated colour-coded masks that show the shadow cone from the WD companion (green), and the tail where the first detonation converges at the surface of the exploding WD (blue), versus the regular ejecta (red). For each of the three series of figures listed above, the masks are generated in the same way as the maps: a slice (not time-dependent, since purely geometrical), a sum in projection weighted by the quantity of interest (slightly time-dependent, since the quantity may be evolving), the intersection with the wave fronts (slowly time-dependent, as the waves are moving). 
For the slice and projection maps, on each figure we show the results along three directions (the three principal axes x, y, z of our simulation box), so as to grasp the entire morphology of the SNR. By design the spherical maps show the entire surface of the SNR at once.

\paragraph{Slices} 
Slices of the mass density are shown in Figure~\ref{fig:map_cut_rho} at select times, and a movie from 1~yr to $4\,000$~yr by steps of 1~yr is available online. The projected position of the companion WD, assuming ballistic motion, is indicated by a white cross, it is still inside the SNR by the end of the simulation. Over time, the characteristic shell structure develops, bounded by the RS and FS. The RTI grows on top of the irregular ejecta boundary. While the instability is in the linear phase, the fingers grow exponentially in time, the resolution of our simulation does not allow us to probe this phase in detail. According to \cite{Schulreich2022RTIeROSITAbubbles}, the instability saturates, and enters the non-linear phase, by the time the extent of the perturbations has reached half their wavelength, which in our simulation happens within the first few years of the SNR evolution. For hundreds of years after, we see that the evolution of the RTI fingers is essentially self-similar. After about a thousand years, we see that the fingers start merging with one another.

Looking at features from the explosion, we note the persistence of a dense region in the centre of the remnant. The shell is off-centre, as a result of the initial velocity shift -- so the geometric centre of the remnant is not the explosion point. There are two other prominent features that stem from the explosion. First, the tail from the first detonation is visible past the average shock surface. At early times (around a hundred years after the explosion) it looks like a protrusion, being ahead of the average FS surface, but actually it is present from the beginning, and is being overtaken by the bulk of the expanding ejecta. This feature is mostly erased after a few hundred years. Second, the conical shadow from the companion WD is visible as a deformed RS. In this under-dense region the RS is travelling faster, the net result after a few hundred years is that the shock front looks segmented, with a straight edge across the shadow. Also we note that the edges of the cone have enhanced RT fingers growth.

The RS moves inward inside the ejecta,\footnote{For a distant observer, the motion of the RS is first outward and later inward, as the RS velocity overtakes the SNR expansion velocity. Regardless, the RS is continuously sweeping up additional mass and thus moving to lower mass coordinates.} reaching the centre at about t = $2\,690$~yr. This time is expected to scale as Eq.~(\ref{eq:t_ch}), that is $E_\mathrm{SN}^{-1/2} M_\mathrm{ej}^{5/6} \rho_\mathrm{ISM}^{-1/3}$. In particular a lower/higher ambient density would mean a slower/faster evolution. After the RS has converged at the centre, there is evidence of a rebound. This feature, which is robust in 3D \citep{Petruk2021Magnetophase}, will be studied in more detail in future work. The peculiar shape of the RS described above is visible all the way from its birth to its disappearance.

\def\widthmapstwo{0.95\textwidth}

\begin{figure}[t!]
\centering
\includegraphics[width=\widthmapstwo]{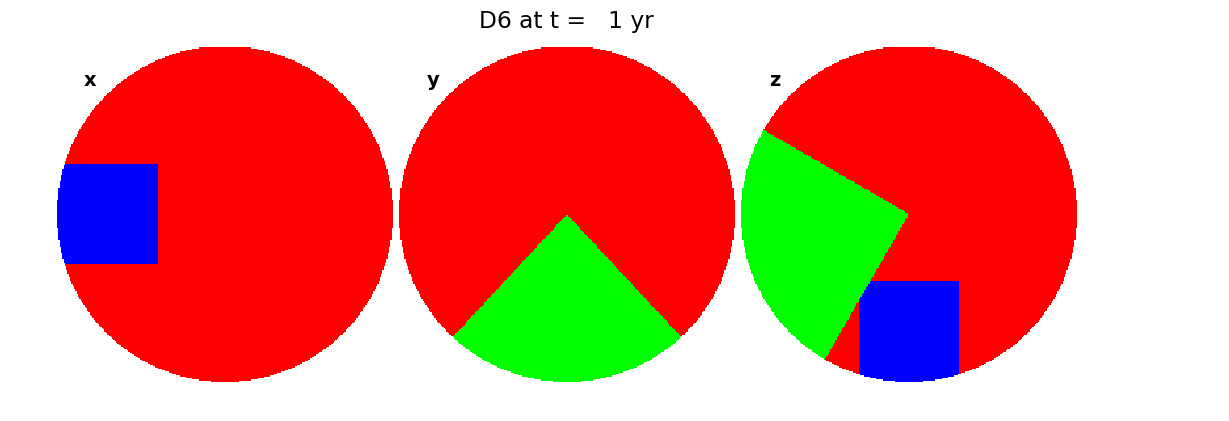}
\includegraphics[width=\widthmapstwo]{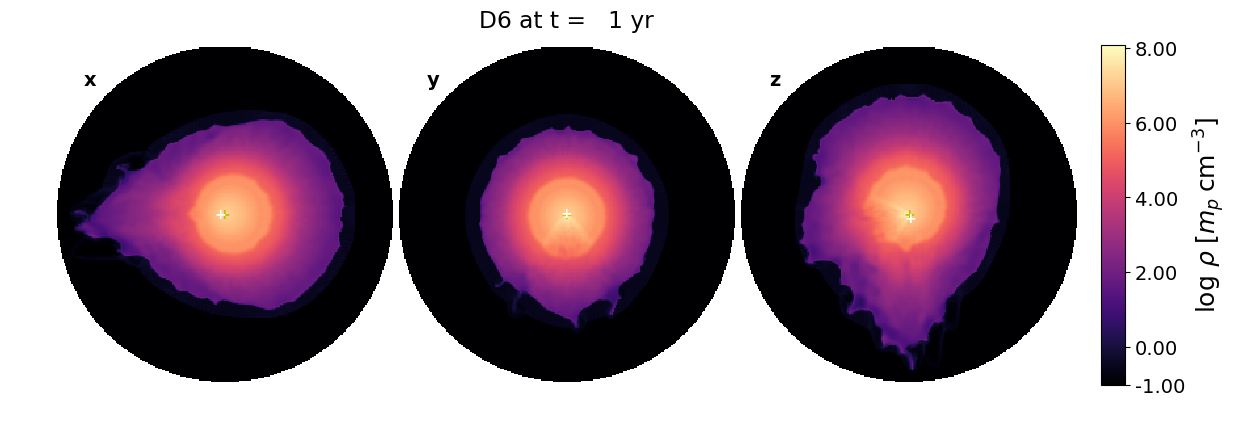}
\includegraphics[width=\widthmapstwo]{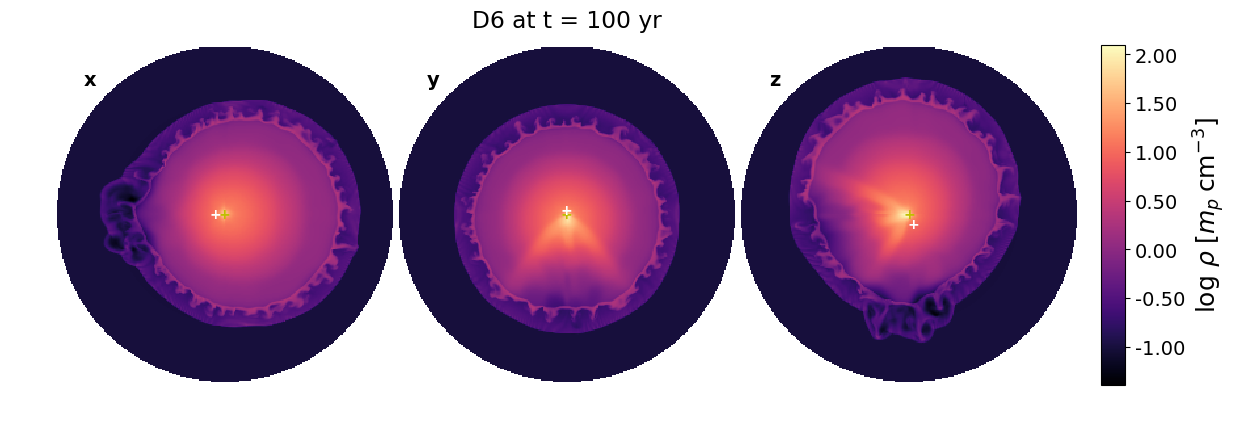}
\includegraphics[width=\widthmapstwo]{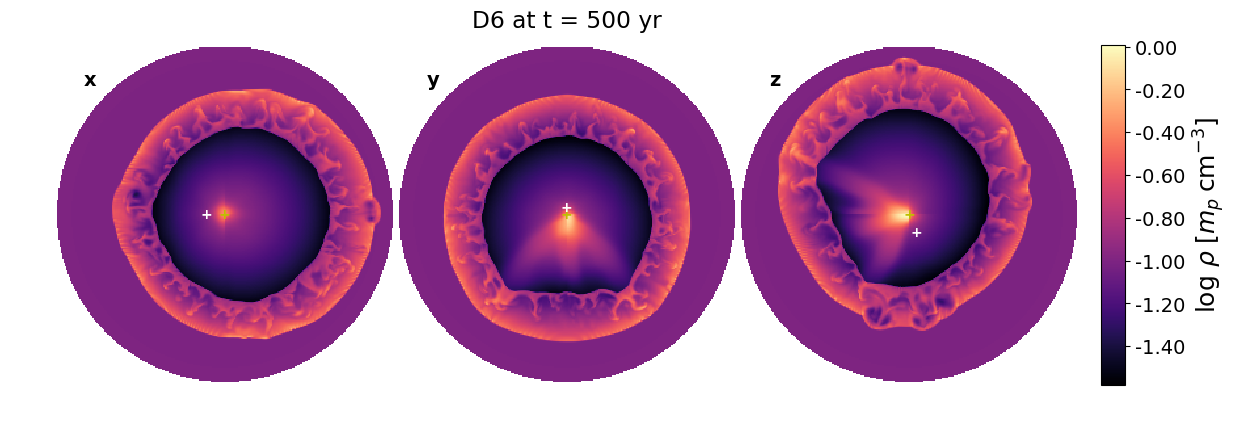}
\caption{Slices of mass density at several times: 1~yr, 100~yr, 500~yr, $1\,000$~yr, $2\,000$~yr, $4\,000$~yr after the explosion, along three different axes (principal axes x, y, z, of the simulation box). 
The box size is 0.14~pc at 1~yr, 5.6~pc at 100~yr, 15~pc at 500~yr, 22~pc at $1\,000$~yr, 30~pc at $2\,000$~yr, 36~pc at $3\,000$~yr, 40~pc at $4\,000$~yr.
Note that the colour scale is shared by the three slices, but adjusted independently at each time.
The yellow cross marks the centre of the exploded WD, the white moving cross marks the position of the surviving WD assuming it keeps its initial velocity (the WD is not resolved).
An animated version of this figure is available online, showing the evolution from 1~yr to $4\,000$~yr by steps of 1~yr (the duration is 1~mn 20~s).
The first plot is a colour-coded mask to aid in identifying the origin of the features: \textcolor{red}{red} = regular ejecta, \textcolor{green}{green} cone = shadow from the companion, \textcolor{blue}{blue} cylinder = tail from the first detonation. These masks are only shown at t = 1~yr since they are not evolving over time.
\label{fig:map_cut_rho}}
\end{figure}
\begin{figure}[t!]
\addtocounter{figure}{-1}
\renewcommand{\thefigure}{\arabic{figure} (continued)}
\centering
\includegraphics[width=\widthmapstwo]{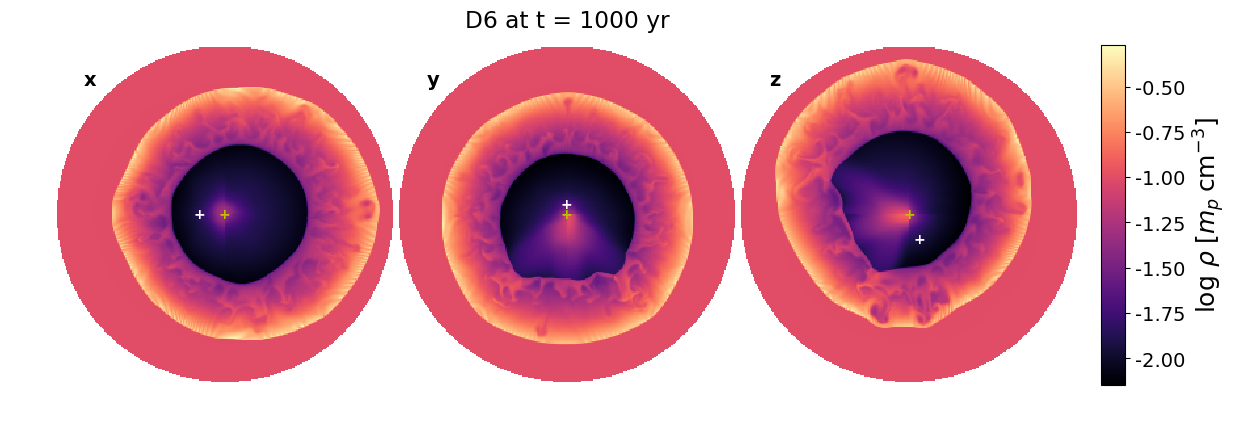}
\includegraphics[width=\widthmapstwo]{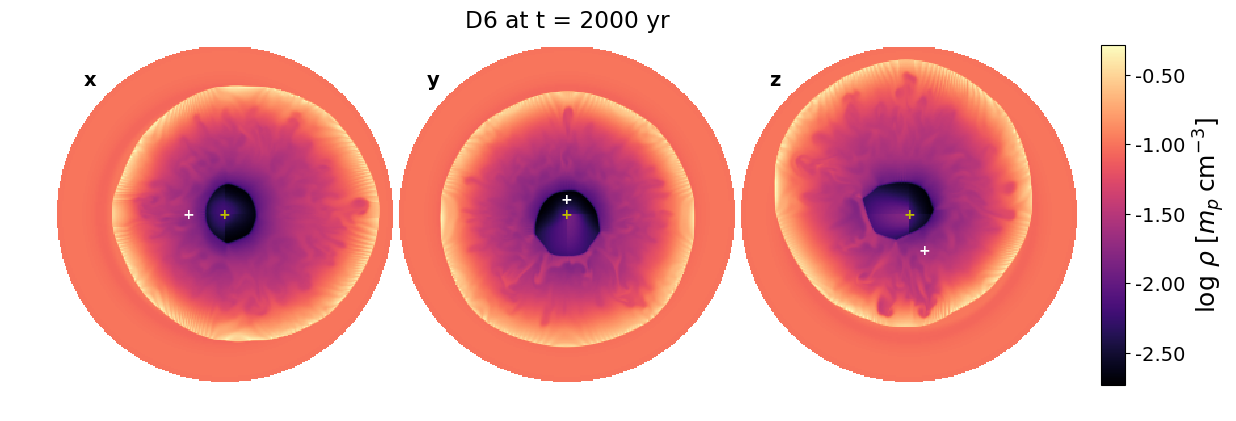}
\includegraphics[width=\widthmapstwo]{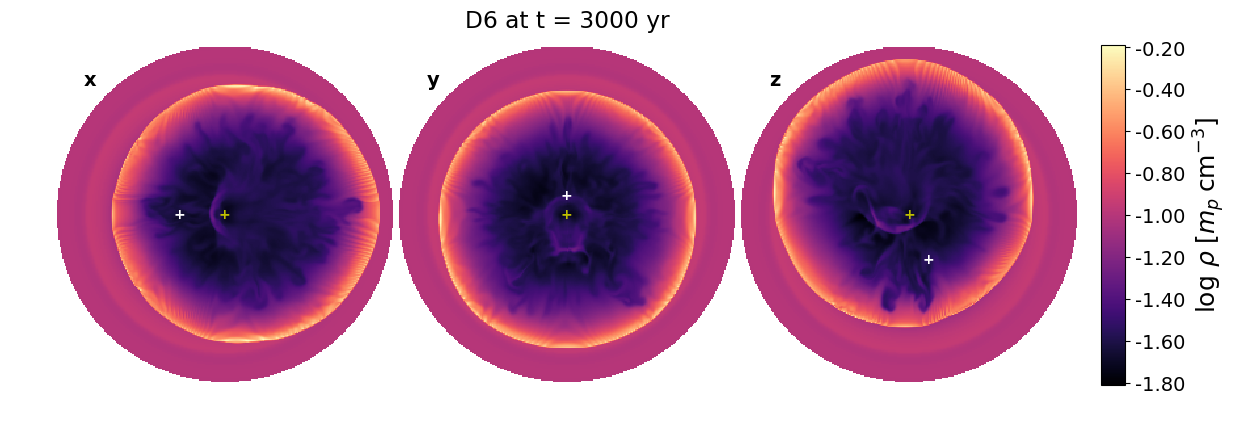}
\includegraphics[width=\widthmapstwo]{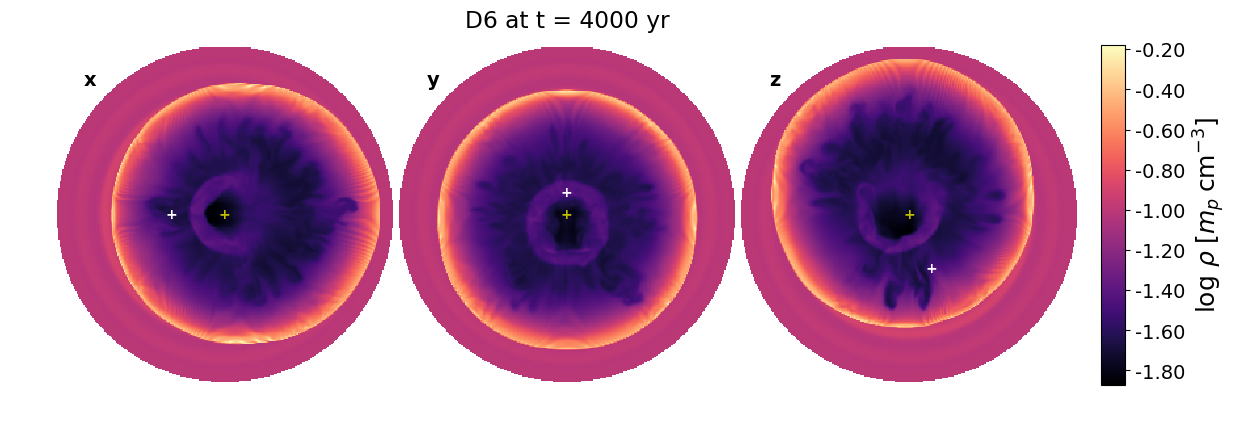}
\caption{}
\renewcommand{\thefigure}{\arabic{figure}}
\vspace{20mm}
\end{figure}

\begin{figure}[t!]
\centering
\includegraphics[width=\widthmapstwo]{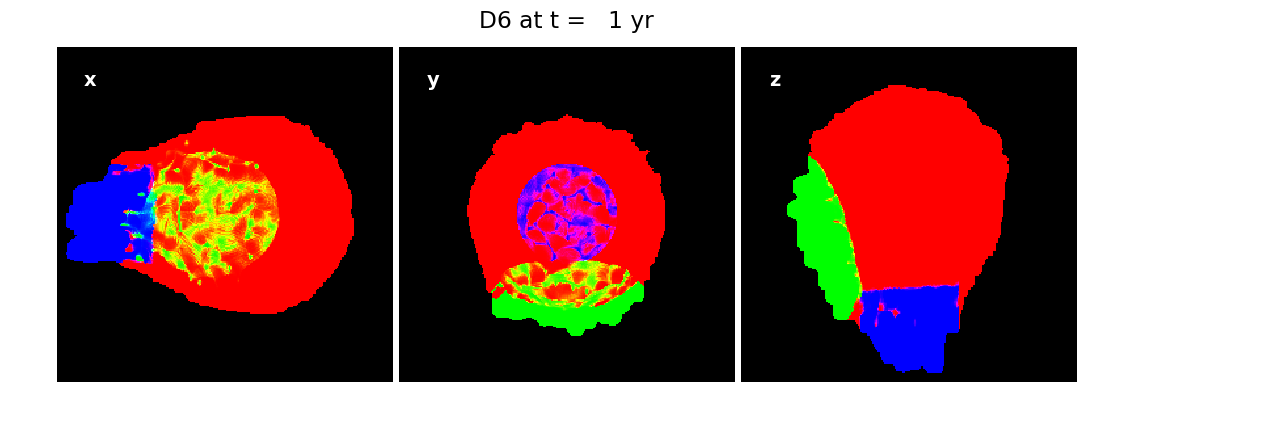}
\includegraphics[width=\widthmapstwo]{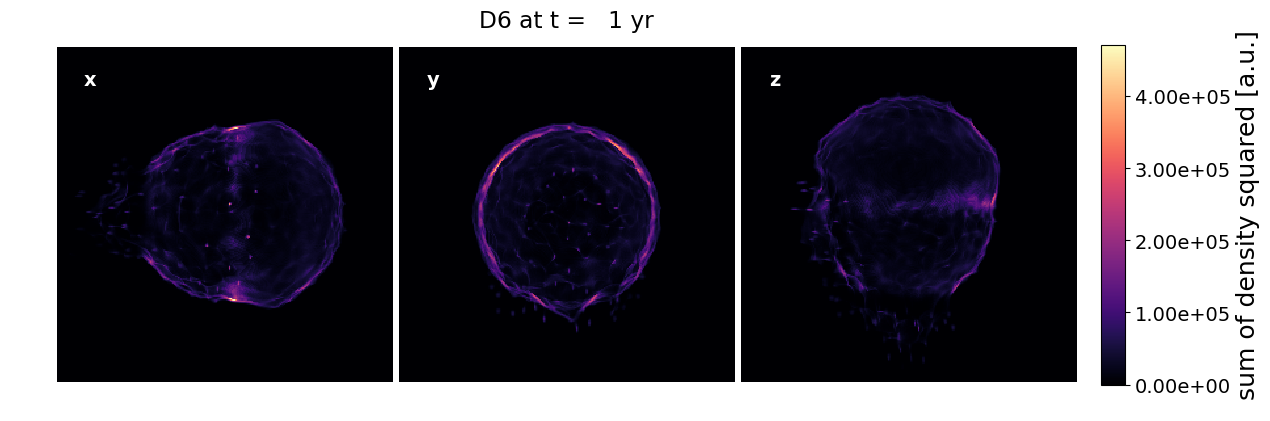}
\includegraphics[width=\widthmapstwo]{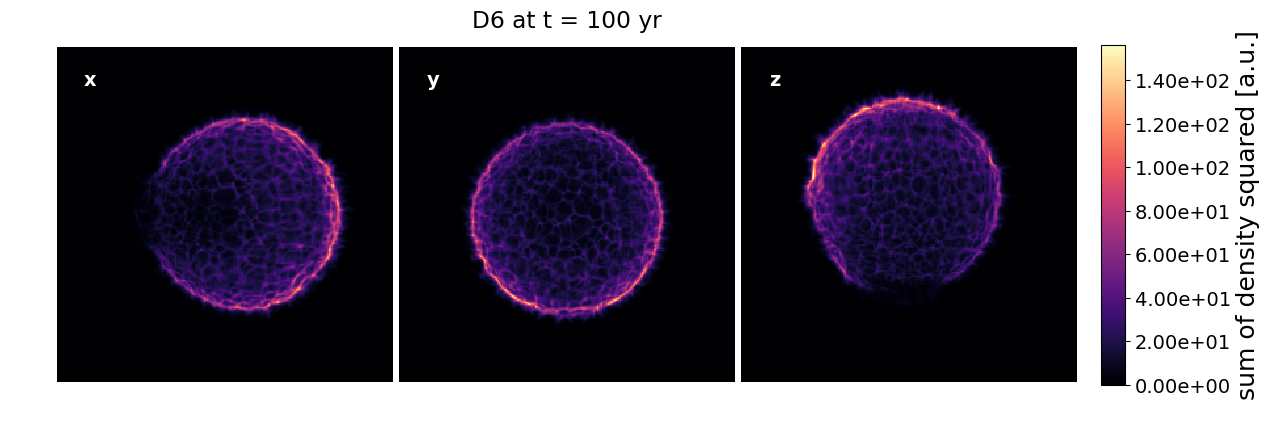}
\includegraphics[width=\widthmapstwo]{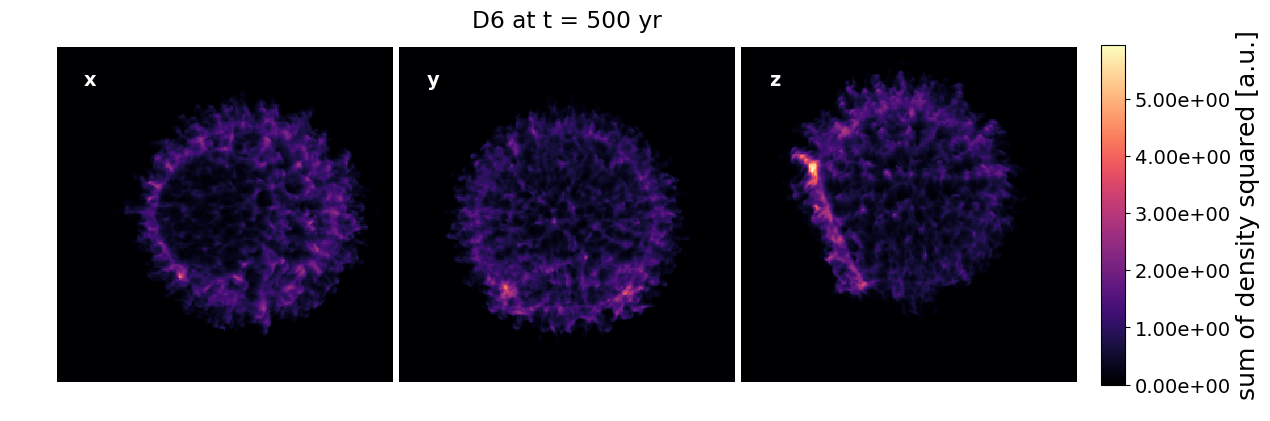}
\caption{Projection of the mass density squared of the shocked ejecta at several times: 1~yr, 100~yr, 500~yr, $1\,000$~yr, $2\,000$~yr, $4\,000$~yr after the explosion, along three different axes (principal axes x, y, z, of the simulation box). The box size at each time is the same as in the previous figure. 
Note that the colour scale is shared by the three slices, but is adjusted independently at each time. 
An animated version of this figure is available online, showing the evolution from 1~yr to $4\,000$~yr by steps of 1~yr (the duration is 1~mn 20~s).
The first plot is a colour-coded mask to aid in identifying the origin of the features: \textcolor{red}{red} = regular ejecta, \textcolor{green}{green} cone = shadow from the companion, \textcolor{blue}{blue} cylinder = tail from the first detonation. These masks are only shown at t = 1~yr since they are mildly evolving over time.
\label{fig:map_prj_f}}
\end{figure}
\begin{figure}[t!]
\addtocounter{figure}{-1}
\renewcommand{\thefigure}{\arabic{figure} (continued)}
\centering
\includegraphics[width=\widthmapstwo]{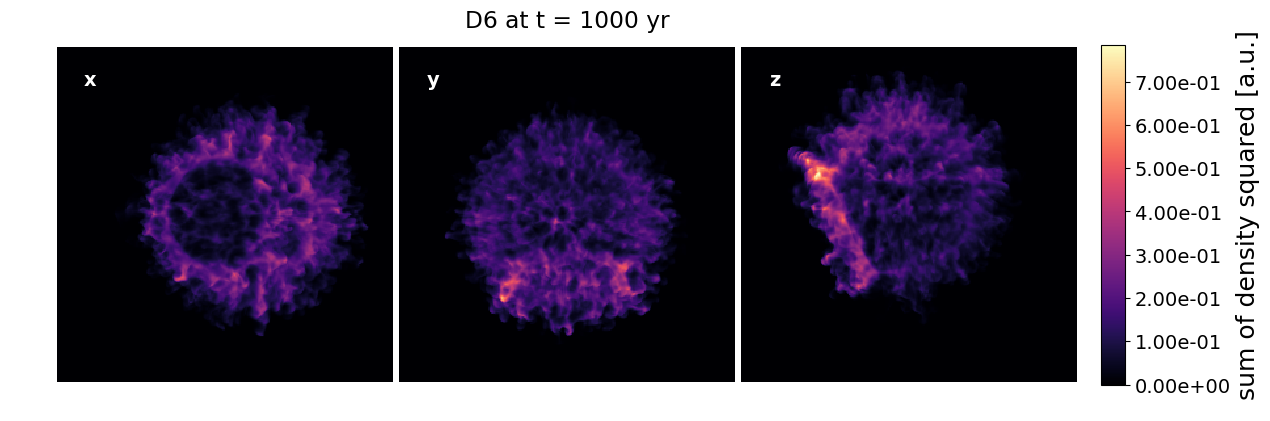}
\includegraphics[width=\widthmapstwo]{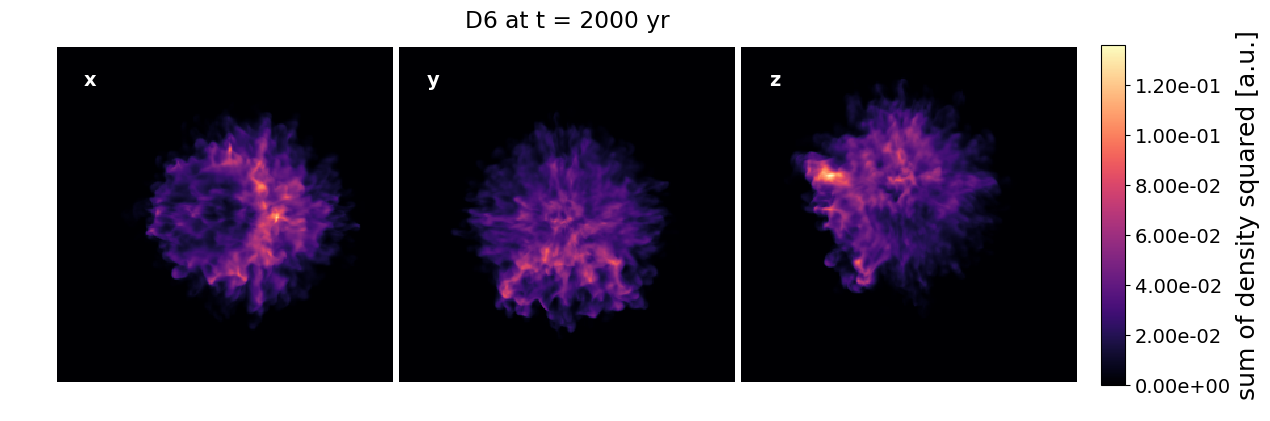}
\includegraphics[width=\widthmapstwo]{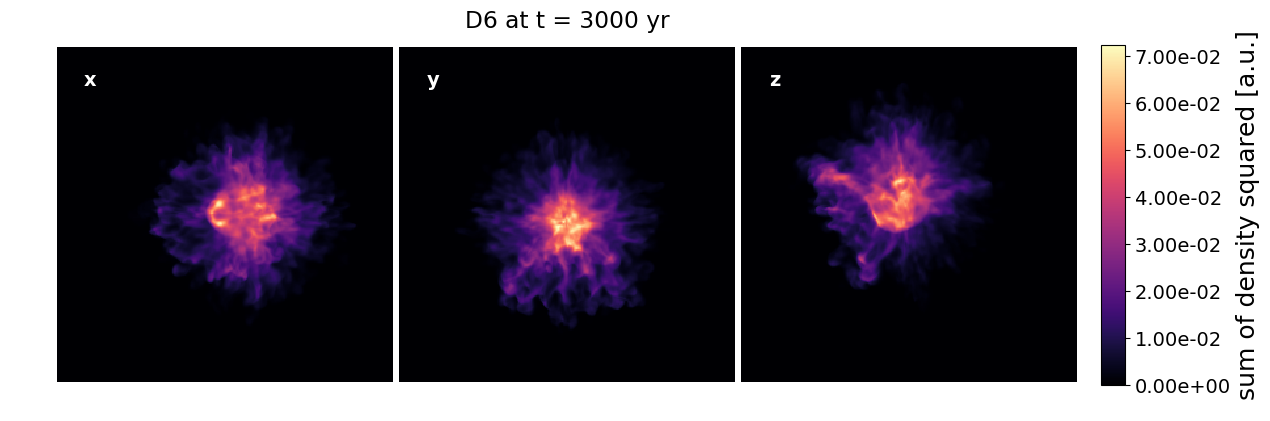}
\includegraphics[width=\widthmapstwo]{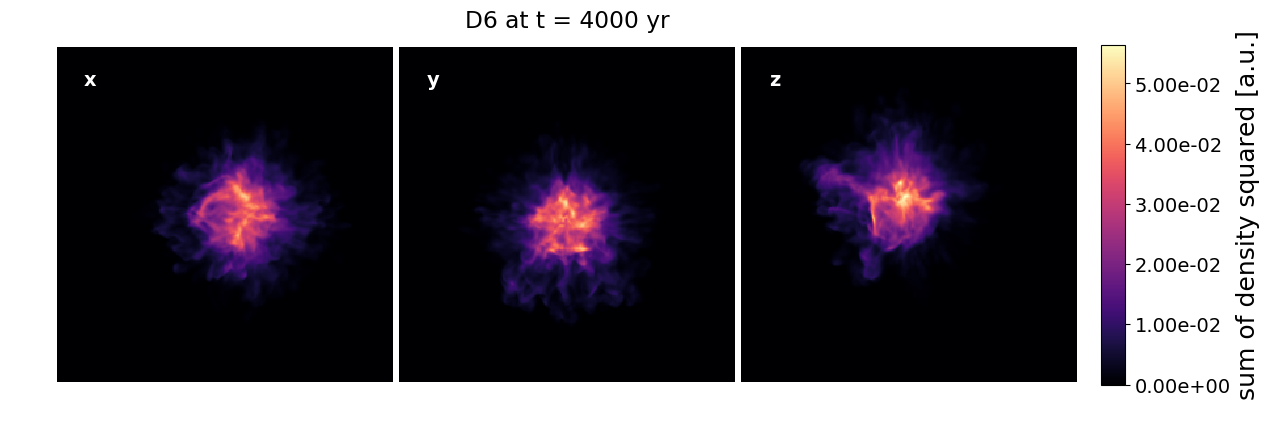}
\caption{}
\renewcommand{\thefigure}{\arabic{figure}}
\vspace{30mm}
\end{figure}

\paragraph{Projections}
Projections of the density squared of the shocked ejecta, a proxy for their broad-band thermal X-ray emission, are shown in Figure~\ref{fig:map_prj_f} at select times, and a movie from 1~yr to $4\,000$~yr by steps of 1~yr is available online.
Most of the small-scale structures visible on these maps come from the RTI. Over time the distance between the CD and the FS is increasing, and so in a comoving frame the ejecta appear to be shrinking in size. Only the shocked ejecta are shown, and so the central density peak is visible only when the RS reaches the centre of the SNR. 
In projection, the tail from the first detonation is more difficult to see. The shadow from the companion is well visible: it appears as a dark disk surrounded by a brighter ring. The dark disk corresponds to an under-dense region, while the bright ring corresponds to enhanced RTI. These features are actually getting more marked over time. They are visible until the rebound of the RS. Since the shadow is a localized feature, the SNR morphology is different depending on the direction of observation. Along the x-axis one is looking almost along the conical axis and so one sees the entire darker disk, along the y-axis one is looking at some angle and sees a bright ellipse, and along the z-axis one is looking sideways and sees a bright bar on the edge.

Initially the shocked ejecta dominate the overall shocked mass and thus the thermal X-ray emission, but eventually the shocked ISM dominates in mass.\footnote{For reference, the time at which the swept-up ISM mass is equal to the ejected mass, called the Sedov-Taylor time, is $t_{\rm ST} = 465$~yr for our assumed parameters. The SNR dynamics will follow the Sedov-Taylor solution for $t \gg t_{\rm ST}$.} When adding the contribution for the shocked ISM using the same proxy of density squared summed along the line of sight, the D$^6$ signature features are still visible at 500~yr, but become hidden at about $1\,000$~yr. So one needs to separate the emission of the shocked ejecta and of the shocked ISM, which is possible using spatially-resolved X-ray spectroscopy, since they have different composition and thermodynamic state. In a subsequent paper focused on the thermal X-ray emission, we will present the separate contributions of the different media and of the different elements they contain. Furthermore, when particle acceleration happens, the induced non-thermal (synchrotron) emission traces the outline of the FS at X-ray wavelengths.

\paragraph{Wave fronts} 
The SNR shell is bounded by two shocks, the RS and the FS, while the ejecta and the ISM are separated by one interface, the CD. Spherical maps and angular spectra of the location of CD, RS, FS are shown in respectively Figures~\ref{fig:healpix_CD}, \ref{fig:healpix_RS}, \ref{fig:healpix_FS} at select times, and movies from 1~yr to $4\,000$~yr by steps of 1~yr are available online.\footnote{Note that the RS is not picked up by the shock tracking system during its final collapse to the centre and after its rebound, even though it is visible in the slices and projections. Angular spectra are no longer reliable when this happens.}
The spherical maps allow us to see at a glance the entire surface considered. The angular spectra, as a function of angular wavenumber $\ell$, allow us to quantify which angular scales contribute to the overall morphology of the SNR. 
For each wave front we observe SN modes at large scales (small $\ell$), that decay quickly in time, except for a permanent dipole ($\ell=1$, meaning two-sided) which comes from the velocity shift from the binary motion. For the CD, we see that the RTI is growing from the smallest scales (largest $\ell$) to larger scales (smaller $\ell$), as expected. In the first part of the simulation, the RTI spectral distribution is fairly symmetric, with a central peak slowly shifting to lower~$\ell$. After about a thousand years, the distribution gets increasingly skewed toward lower~$\ell$, as fingers growth is now driven by mergers.
Maps for the RS and the FS are simpler than the ones for the CD, although they bear some imprint of the RTI as well: the feet of the fingers for the RS, the tips of the fingers for the FS.
The tail from the first detonation is visible at the beginning. Then the ring from the companion shadow appears, seen on the spectra in the low-$\ell$ modes. 

\def\widthhealpix{1\textwidth}

\begin{figure}[t!]
\centering
\includegraphics[width=\widthhealpix]{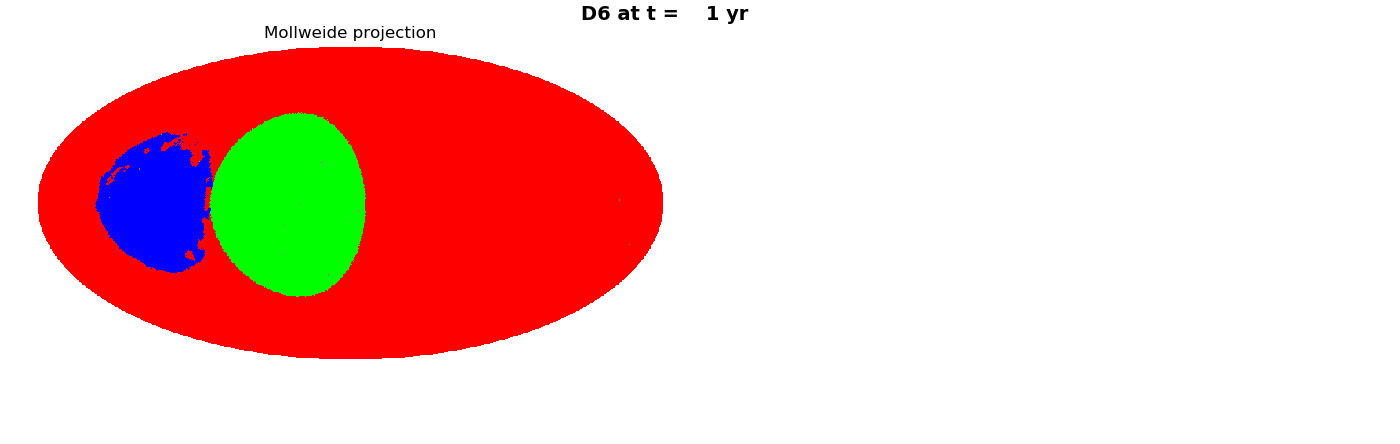}
\includegraphics[width=\widthhealpix]{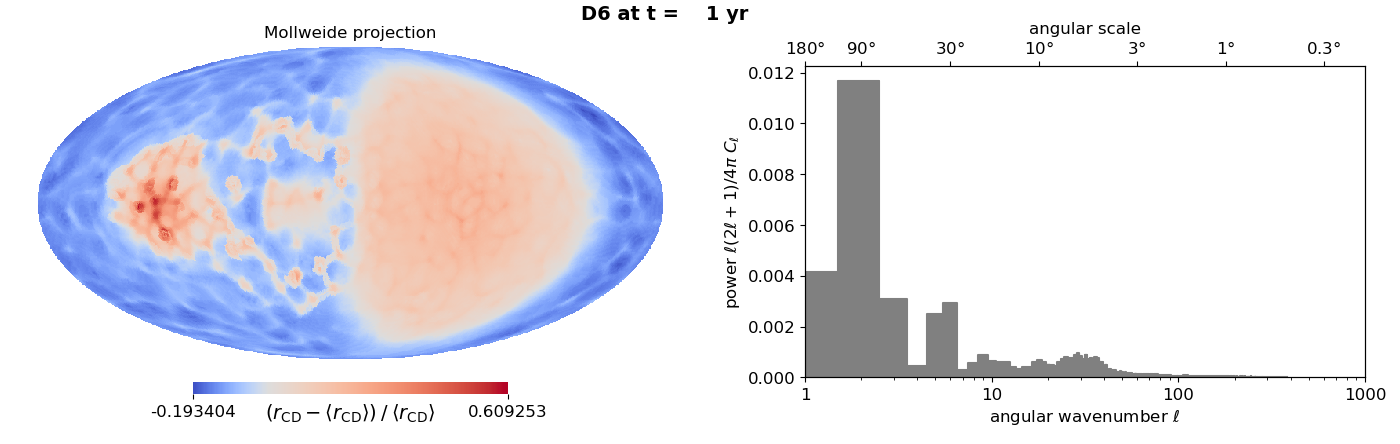}
\includegraphics[width=\widthhealpix]{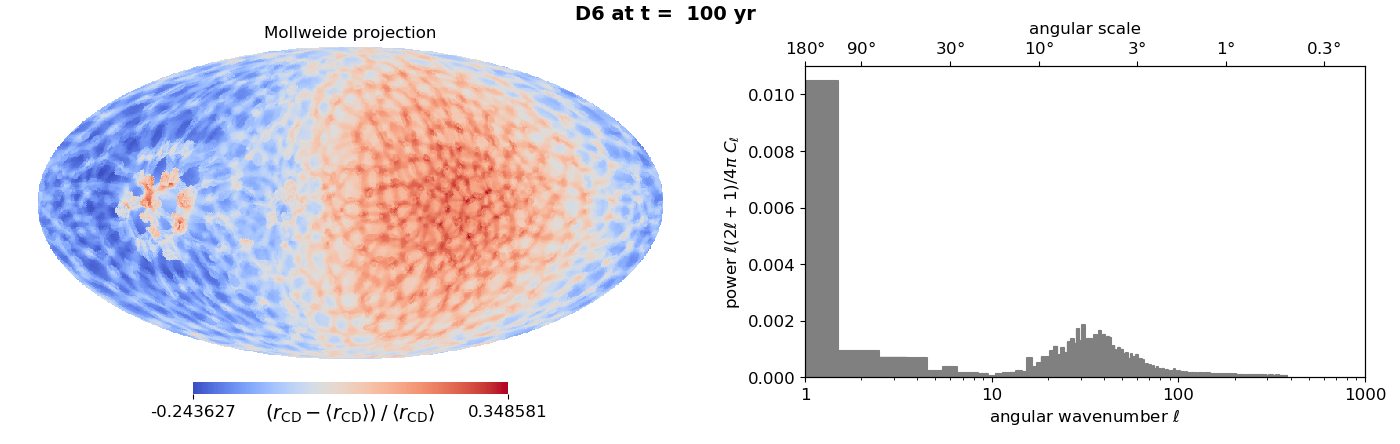}
\includegraphics[width=\widthhealpix]{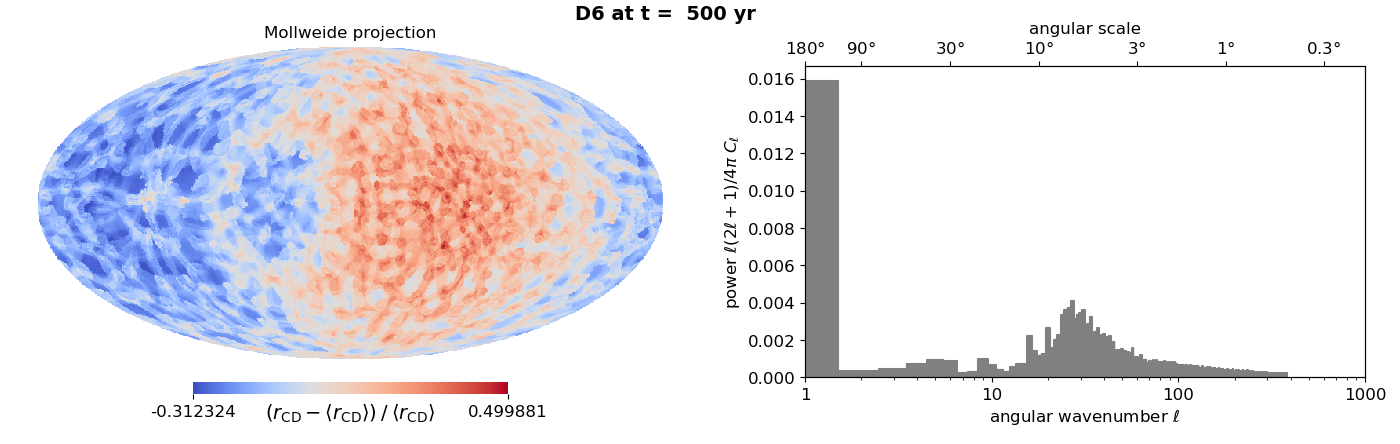}
\caption{Morphology of the contact discontinuity. 
Maps on the left are spherical projections of the radial variations of the location of the wave. We use the Mollweide projection; for all times it is centred on the dipole component at the initial time.
Spectra on the right result from an expansion in spherical harmonics of these variations. At angular wavenumber~$\ell$, the typical angular scale probed is $\pi/\ell$, and the power $C_{\ell}$ plotted is normalized in such a way that each grayed bin is the contribution of wavenumber~$\ell$ to the total variance of the radial fluctuations.
Three times are shown: 1~yr, 100~yr, 500~yr. The maps and spectra evolve slowly after that.
An animated version of this figure is available online, showing the evolution from 1~yr to $4\,000$~yr by steps of 1~yr (the duration is 1~mn 20~s).
The first plot is a colour-coded mask to aid in identifying the origin of the features: \textcolor{red}{red} = regular ejecta, \textcolor{green}{green} cone = shadow from the companion, \textcolor{blue}{blue} cylinder = tail from the first detonation. These masks are shown at t = 1~yr.
\label{fig:healpix_CD}}
\end{figure}

\begin{figure}[t!]
\centering
\includegraphics[width=\widthhealpix]{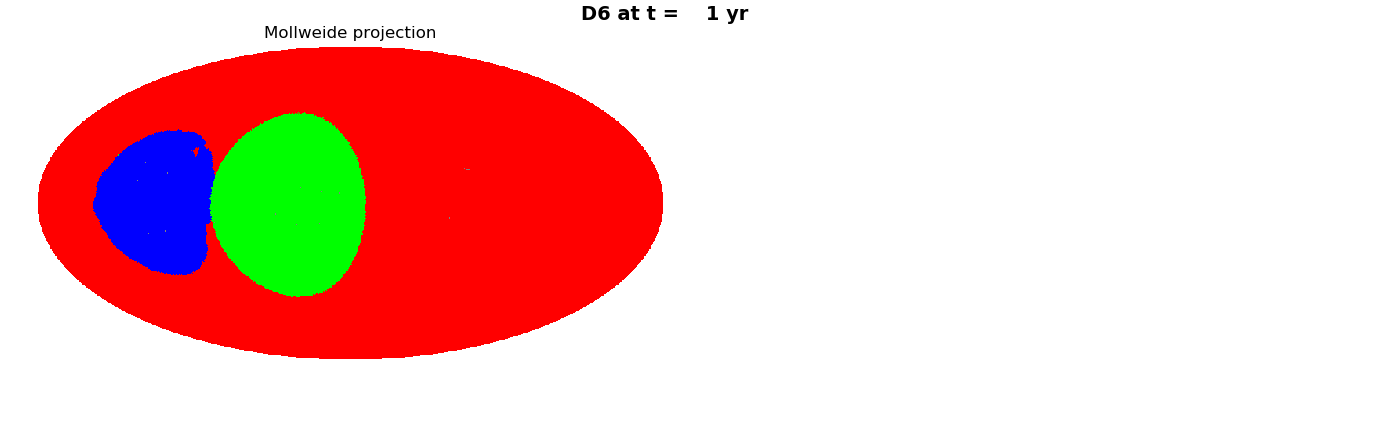}
\includegraphics[width=\widthhealpix]{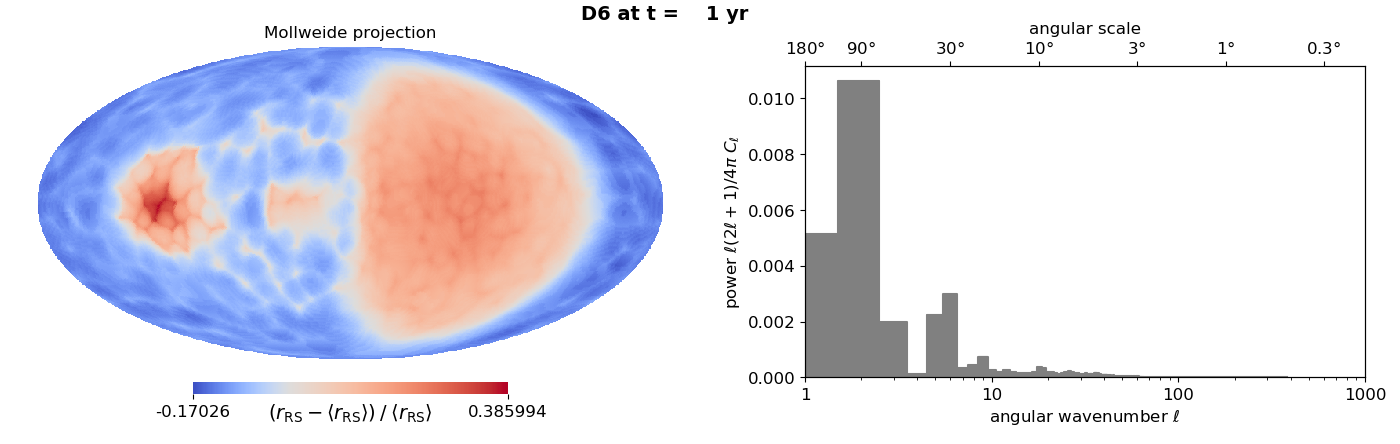}
\includegraphics[width=\widthhealpix]{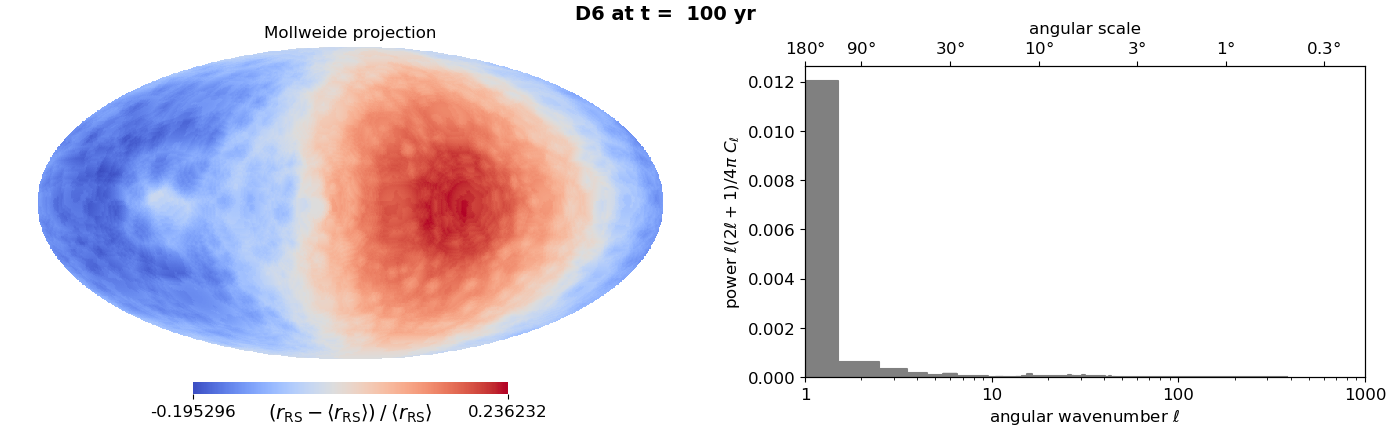}
\includegraphics[width=\widthhealpix]{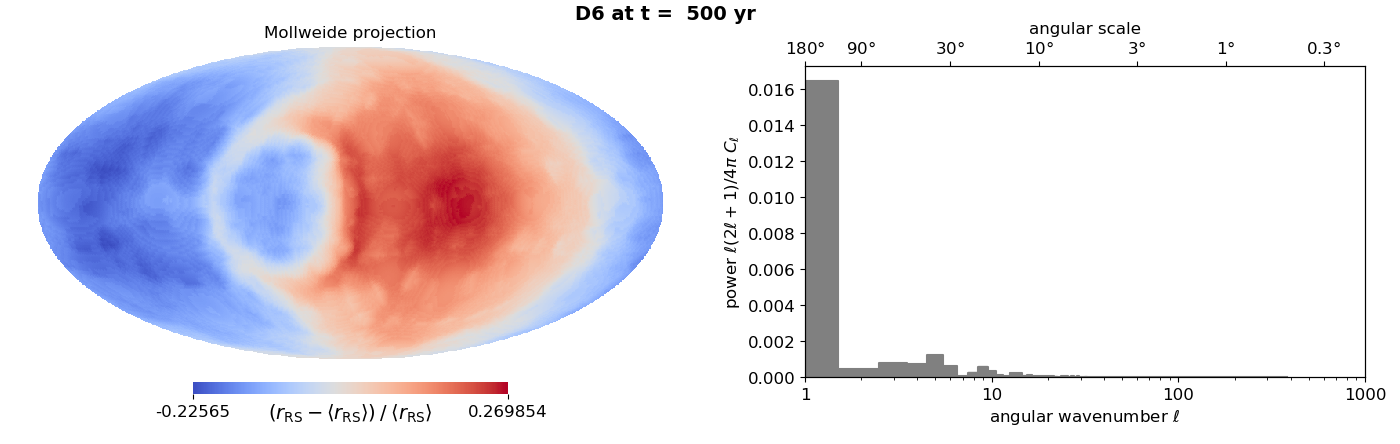}
\caption{Morphology of the reverse shock. 
Same as Figure~\ref{fig:healpix_CD}, for the radial variations of the RS location.
Three times are shown: 1~yr, 100~yr, 500~yr. An animated version of this figure is available online, showing the evolution from 1~yr to $4\,000$~yr by steps of 1~yr (the duration is 1~mn 20~s).
\label{fig:healpix_RS}}
\vspace{20mm}
\end{figure}

\begin{figure}[t!]
\centering
\includegraphics[width=\widthhealpix]{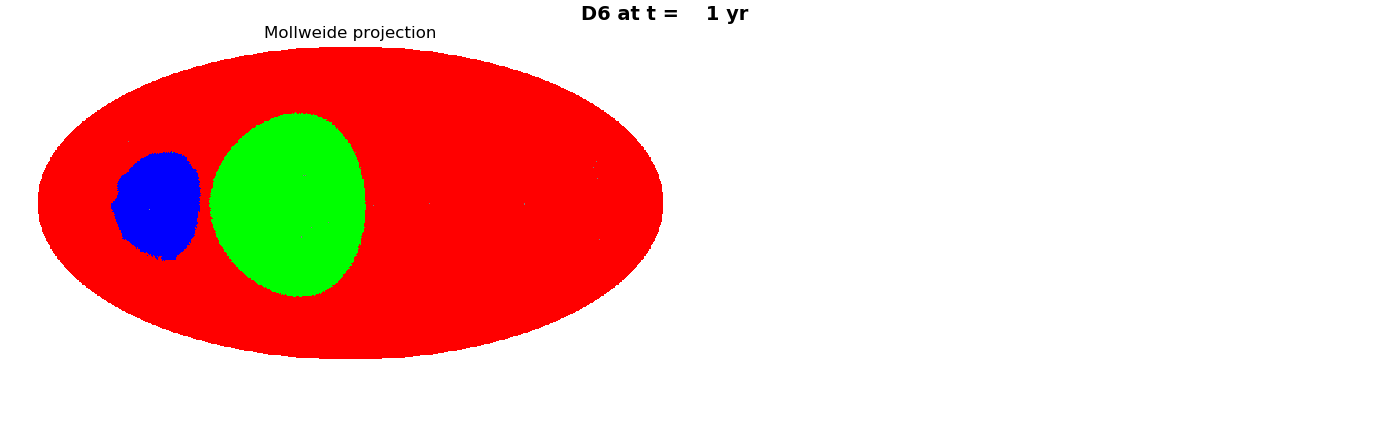}
\includegraphics[width=\widthhealpix]{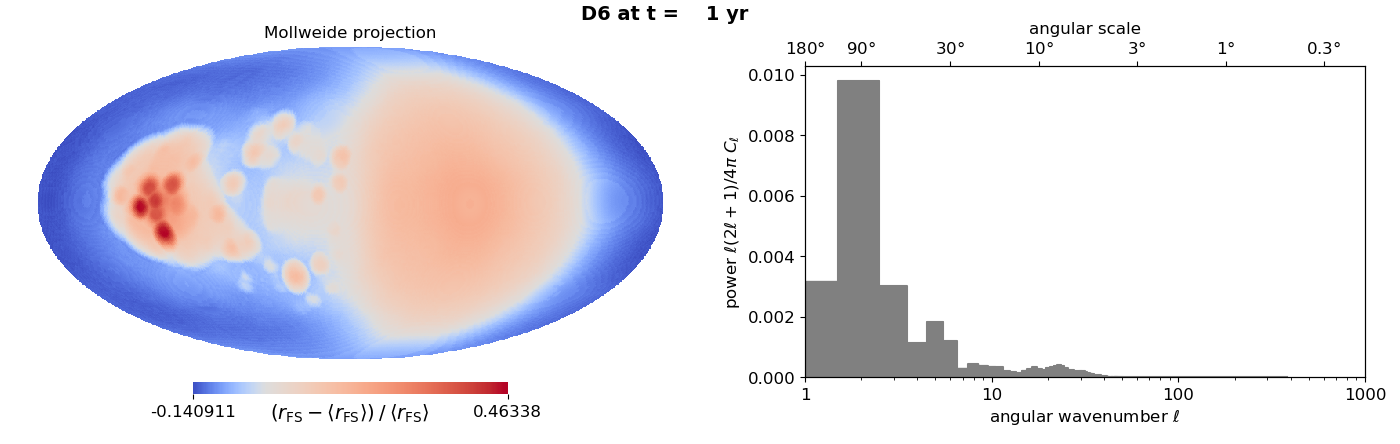}
\includegraphics[width=\widthhealpix]{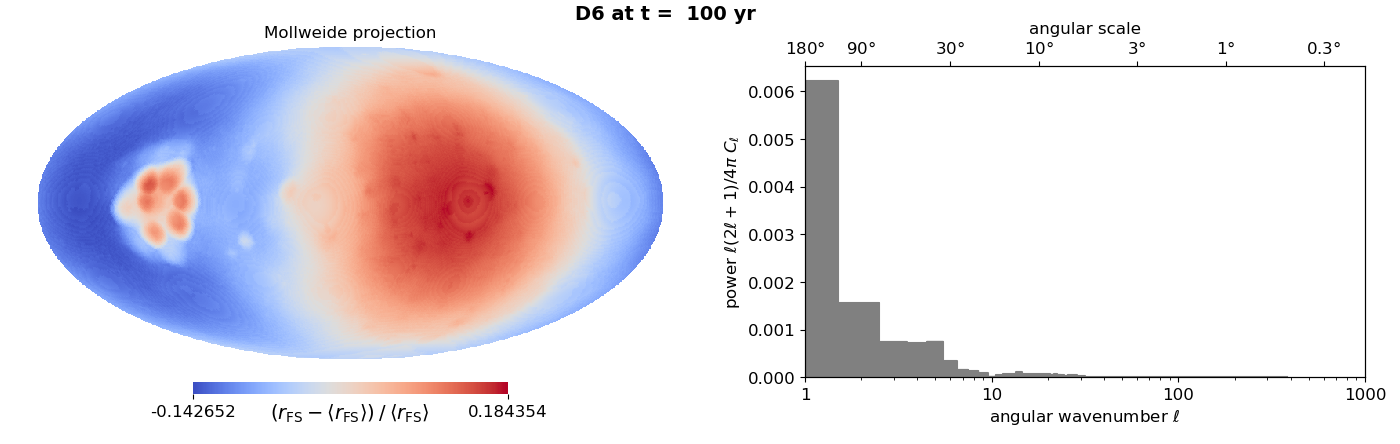}
\includegraphics[width=\widthhealpix]{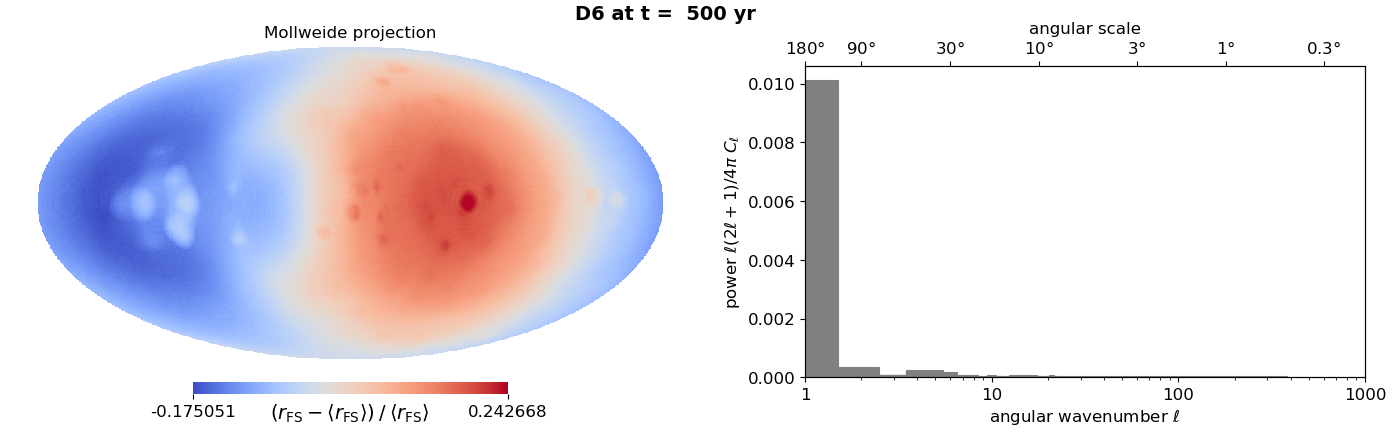}
\caption{Morphology of the forward shock. 
Same as Figure~\ref{fig:healpix_CD}, for the radial variations of the FS location.
Three times are shown: 1~yr, 100~yr, 500~yr. An animated version of this figure is available online, showing the evolution from 1~yr to $4\,000$~yr by steps of 1~yr (the duration is 1~mn 20~s).
\label{fig:healpix_FS}}
\vspace{20mm}
\end{figure}

\def\widthpower{0.9\textwidth}
\def\heightpower{0.6\textwidth}
\begin{figure}[t!]
\centering
\includegraphics[width=\widthpower,height=\heightpower]{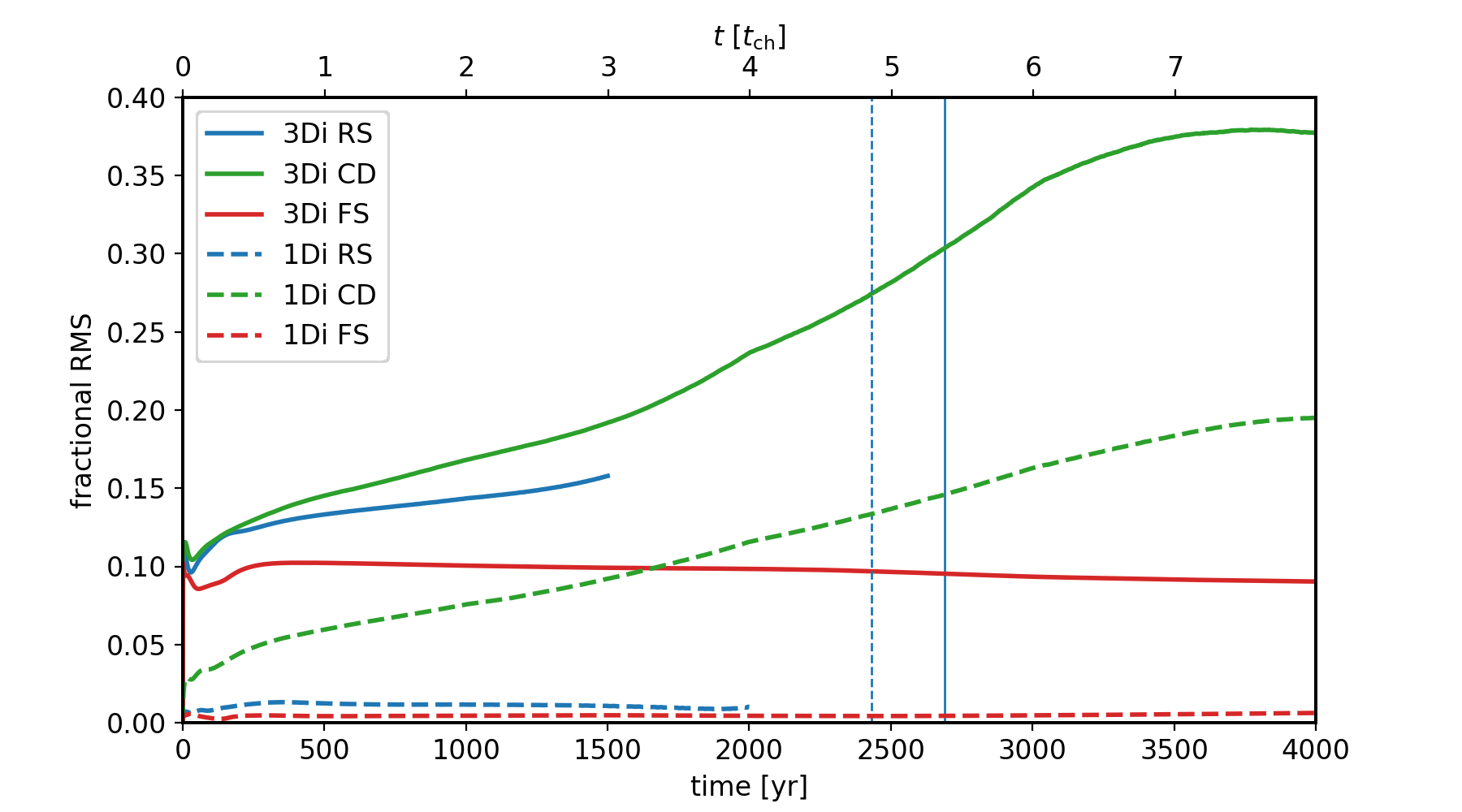}
\includegraphics[width=\widthpower,height=\heightpower]{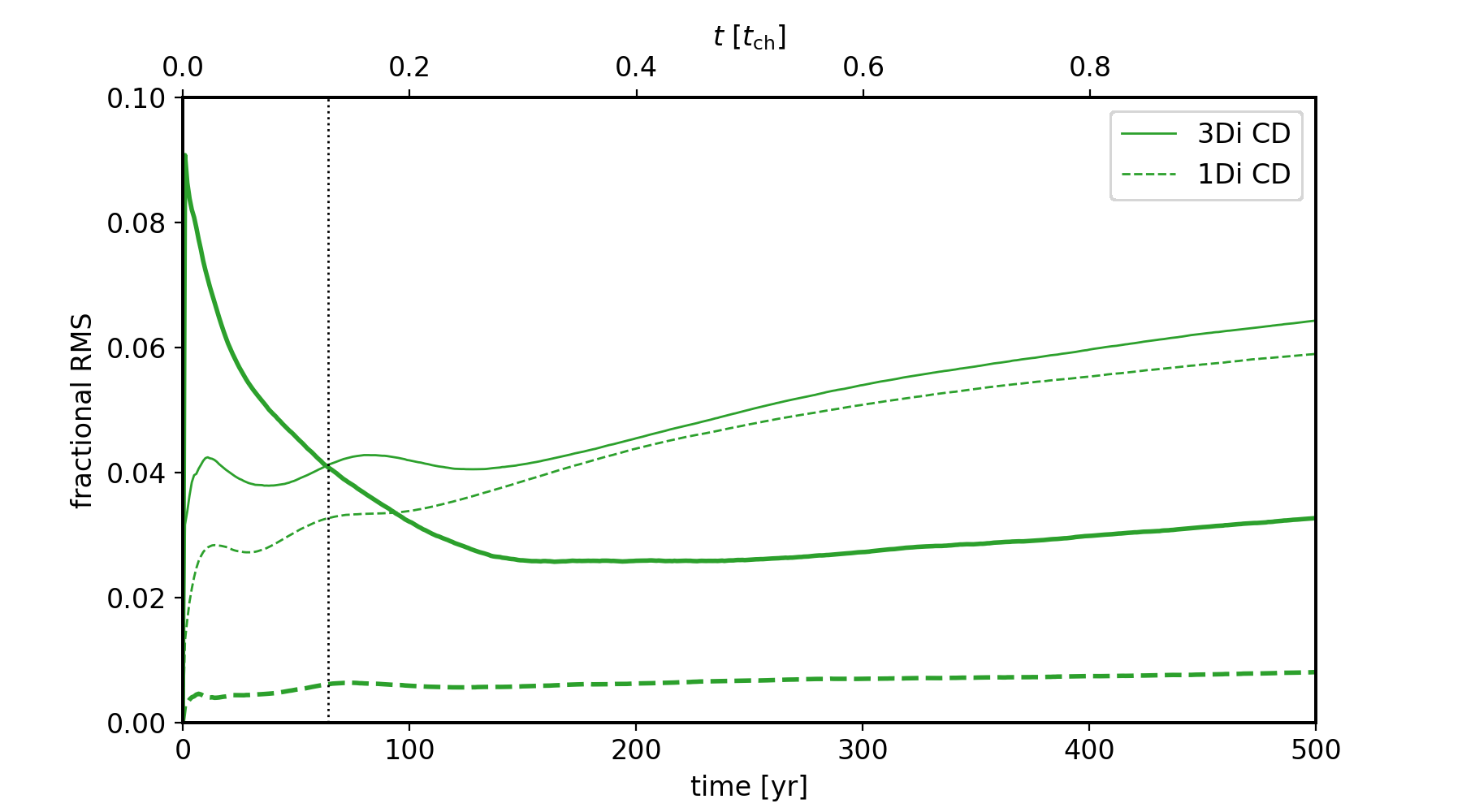}
\caption{Fractional RMS (square-root of the angular power) as a function of time, for the three wave fronts: FS in red, CD in green, RS in blue. 
Time is indicated in years and in characteristic timescale as defined by equation~(\ref{eq:t_ch}). 
Two cases are compared: spherically symmetric ejecta (1Di, dashed curves) versus asymmetric ejecta (3Di, solid lines).
The top panel shows the evolution of the total power, for each interface, until the final simulation time of $4\,000$~yr. The curves for the RS are stopped when the tracking of its surface, and therefore the calculated angular power, becomes unreliable. The time when the RS reaches the centre is indicated by the vertical blue line. 
The bottom panel shows the power separately at large scales, excluding the dipole ($1<\ell<10$, thick lines) and at small scales ($\ell\geq10$, thin lines), only for the CD, and up to 500~yr after the explosion. The vertical dotted line indicates the time at which the two curves intersect for the 3Di simulation. 
\label{fig:power}}
\end{figure}

As a final step, we aim to quantify which of the initial conditions (progenitor system and explosion mechanism) and of the ejecta-ISM interaction (growth of the RTI) shape the SNR at a given age. The summed angular power in the wave front deformations is shown in Figure~\ref{fig:power} as a function of time. The top panel shows the evolution until the simulation end time $4\,000$~yr, for each of CD, RS, FS, for both 1Di and 3Di cases. 
The 1Di case uses smooth initial conditions. For the FS and RS nothing is expected to happen and so this is merely a test of the numerical precision; for the CD this case shows the effect of purely the RTI. The 3Di case uses the actual initial profiles, and shows the imprint of the explosion. On the angular spectra, we see that the SN modes (at large scales) and the SNR modes (at small scales) can be separated for the first few hundred years around $\ell=10$; the power in these two separate $\ell$ bands is plotted in the bottom panel, for the CD only, up to 500~yr. At small scales (excluding the permanent dipole, thin lines) the evolution is similar in the 1Di and 3Di cases, which comforts us in the fact that it corresponds to RTI growth. At large scales (thick lines), in 1Di nothing happens as expected, while in 3Di a decay is visible, strong initially but not reaching zero level. Looking at the 3Di case (solid lines), the low-$\ell$ and high-$\ell$ curves for the CD intersect at 85~yr. It~means that, after that time, most of the angular power in the ejecta morphology is coming from the SNR RTI, which is in its non-linear phase of self-similar growth. However signatures of the explosion are still visible on the maps at much later ages, because they are localized in space, stable in time features. This is in contrast with our previous studies in Papers~I and II with the N series of models, for which signatures were seen as statistical deviations across the entire SNR surface.
\\

\section{Discussion} 
\label{sec:discussion}

We have presented for the first time the evolution and morphology of a D$^6$ SNR, up to $4\,000$~yr after the explosion, which with our chosen ISM density is after all the ejecta have been shocked. In~this section we discuss implications for the interpretation of observations of SNRs.

\subsection{Other effects}

First we comment on aspects of our modelling that may be simplifications of the physical problem.

\paragraph{Effect of a different environment} We have deliberately used a very simple homogeneous ambient medium, in order to study the impact of the initial conditions. Certainly the ISM could be more complicated, especially after a few thousand years of evolution when the blast wave has covered tens of parsec, and that would impact the SNR as well. What we are demonstrating in this work, is that an inhomogeneous ISM is not required in order to produce a complex SNR morphology. The morphology we have been exploring is what is obtained, even in a uniform ISM, from the D$^6$ explosion physics. Type Ia SNRs are known to have more regular shapes overall than core-collapse SNRs \citep{Lopez2011UsingMixing}, but they do not necessarily have symmetric ejecta.

\paragraph{Effect of radioactive decay of $^{56}$Ni} We investigated the effect of heating from radioactive decay with a simplified approach, as in Papers~I and~II.\footnote{Another radioactive element of interest to SN~Ia is $^{44}$Ti. However its yield is expected to be much smaller than that of $^{56}$Ni, and it is not reliably calculated in our SN simulation.}
We observed that heating tends to smooth the features, especially the central density peak, which thus may not be a robust feature. On projected maps, the main features we discussed in the previous section (tails from the first detonation, shadow from the companion) are still visible, just somewhat washed out. We recall that our simplified treatment, with local deposition of the available energy, provides the maximum possible impact of nickel decay. 
We note that centrally bright SNRs are observed: the so-called mixed-morphology SNRs (or thermal composites) that display radio shells with centrally peaked thermal X-rays, but these tend to be older objects and associated with massive star explosions (see the discussion in Paper~II). It would be interesting to confirm whether some objects from this class originate from thermonuclear explosions. 

\subsection{Comparison with previous works}

As mentioned in the introduction, only a few previous works have looked into the signature of a companion star on the SNR, some driven by observations \citep{Lu2011TheCompanion,Vigh2011AsymmetriesRemnants}, others more theoretical in nature \citep{Garcia-Senz2012IsRemnants,Gray2016ShadowsRemnants}. We now compare our results to their findings.

\cite{Garcia-Senz2012IsRemnants} ask whether there is ``a hidden hole'' in Type Ia SNRs. They perform 2D axisymmetric SPH simulations, that are also done in two steps, the SN then the SNR. Using an existing spherical model for the ejecta, they first compute the interaction of the ejecta with the companion, then remove the companion, scale up the ejecta, add a uniform ambient medium, and follow the SNR evolution from 28~yr to $1\,000$~yr. They assume the companion to be a main-sequence solar-like star, and that it makes a ``hole'' in the ejecta up to a half-opening angle $\theta = 20^\circ$, which is half the size of the ejecta shadow in our D$^6$ model.\footnote{Since there is actually matter everywhere, just less dense and more turbulent around the companion, we prefer to use the term ``shadow'' rather than ``hole''.} They note that the hole can actually affect the SNR on larger angular scales. Their main finding is that the hole can remain open several hundred of years, which is consistent with our own findings. They report that RTI is enhanced at the edges of the hole, which we also observe in our simulations. They state that the hole is visible only along some directions, which we (and \citealt{Gray2016ShadowsRemnants} discussed next) disagree with, since in X-rays the entire SNR is observed in projection. In a follow-up paper \citep{Garcia-Senz2019InteractionLaboratory} the study was complemented with 3D simulations plus a laser laboratory experiment.

\cite{Gray2016ShadowsRemnants} investigate the ``shadows'' of Type Ia SNe companions. They perform 3D SPH simulations, also done in two steps, with the SNR evolution calculated from 100~yr (which is quite late for assuming free expansion) to $1\,000$~yr (although results are shown only up to 300~yr). They assume the companion to be a sub-giant star, and find that it makes a shadow with half opening angle of $\theta = 40^{\circ}$, similar to the ejecta shadow in our D$^6$ model. They also find that the shadow is visible for hundreds of years. They compute a proxy for the thermal X-ray emission, and their maps in Figure~10 are to our knowledge the closest to our present work. As in our maps, the shadow is present at all viewing angles, with a different morphology depending on the direction of observation with respect to the shadow conical axis. Looking along the axis, a darker disk is observed on the surface of the remnant, looking across the axis, a straight bright bar is observed at the edge of the remnant, in-between an ellipse is observed, similar to what we obtained for D$^6$ (with a different kind of companion).

\cite{Lu2011TheCompanion} interpret some features seen in the X-ray images of Tycho's SNR as signatures of an (unseen) companion: a prominent arc would be a bow-shock, and a darker region along the edge would be a shadow. The half-opening angle of the feature is about $\theta \simeq 10^\circ$, which is significantly smaller than the shadow of the aforementioned models and of our D$^6$ model, although comparable with the size of the stream of stripped material. The paper offers no detailed modelling of the ejecta--companion interaction. The companion is assumed to be a normal star. Having presented evidence for the existence of a companion, the authors claim that this is evidence for the single degenerate scenario, which we now know is not correct: a~double degenerate explosion like D$^6$ can leave a surviving companion that can leave its imprint on the SNR. 

It is worth noting that despite using different companion stars, in the numerical models discussed above, the angular size of the companion as seen from the explosion centre is similar: in \cite{Garcia-Senz2012IsRemnants}, for a Sun-like companion, $\theta = 19.5^{\circ}$, in \cite{Gray2016ShadowsRemnants}, for more massive stars, $\theta$ ranges in $24^{\circ}-30^{\circ}$, and in our work, for a WD, $\theta = 22^{\circ}$.
The rough uniformity of angular size is a direct consequence of assuming mass transfer from the companion onto the primary. Assuming that the companion fills its Roche lobe at its orbital distance, its angular size is known to be a function of only the mass ratio $q$ of the binary  \citep{Kopal1959Closebinarysystems}, $q=0.6$ in our model. Using \cite{Eggleton1983Aproximationslobes} formula, the angle $\theta=10^{\circ}$ inferred by \cite{Lu2011TheCompanion} would require $q$ to be about 0.06, which seems unreasonably small. We also note that, being non-thermal, the X-ray arc in Tycho is more likely related to particle acceleration at the shock front.

\cite{Vigh2011AsymmetriesRemnants} observe that Tycho's SNR has a two-sided, East/West morphology, like two hemispheres of slightly different radius. One of the possible causes they identify is mass loading from a companion star (other possibilities include an asymmetric wind of the progenitor). To assess this scenario, they perform hydro simulations, 2D axisymmetric and 3D, with varying opening angle and mass excess. They find that the morphology of Tycho can be reproduced with an opening angle of $90^\circ$ (that is, an entire hemisphere) and a mass excess between 0.3 and 0.6~$M_\odot$. They produce mock X-ray maps (their Figures~9 and~11), that are not as realistic looking as ours or the ones of \cite{Gray2016ShadowsRemnants}. In a follow-up study \citep{ Moranchel-Basurto2020SimulatedModel} they perform 3D MHD simulations, and compute the synthetic synchrotron emission in radio. Whether or not it is appropriate to explain Tycho's SNR, this model is unlike a D$^6$ SNR. 

\subsection{Possible target SNRs}

Having described a D$^6$ SNR, it is natural to ask whether such a SNR has been observed or not. The sample of possible targets is quite limited. With current instruments, morphological studies are possible for SNRs located in our galactic neighbourhood: the Milky Way and the Large and Small Magellanic Clouds (LMC and SMC). Plus we should look for SNRs that are dynamically young enough that their morphology be determined by the explosion rather than by the circumstellar medium. In the models investigated in Papers~I and~II the imprint of the explosion was found to typically last for a few hundred years, though for D$^6$ it can last longer. There are only a handful of nearby SNRs known to be the remnants of Type~Ia SNe and less than about a thousand years of age: G1.9+0.3 (about 150~yr, \citealt{Borkowski2013SupernovaG1.9+0.3}); SNR 0509--67.5 (about 400~yr, \citealt{Warren2004Raising050967.5,Rest2005LightCloud}); Kepler's SNR = SN~1604 = G4.5+6.8 (417~yr, \citealt{Reynolds2007AInteraction,Burkey2013X-RaySupernova}), with a mysterious asymmetric morphology that must be giving us insights into the progenitor system \citep[e.g.][]{Kasuga2021SpatiallyRemnant}; the aforementioned Tycho's SNR = SN~1572 = G120.1+1.4 (449~yr, \citealt{Warren2005Cosmic-RayObservations,Yamaguchi2017TheRemnant}); SNR 0519--69.0 (about 450--600~yr, \citealt{Kosenko2010TheStudy,Rest2005LightCloud}); SN 1006 = G327.6+14.6 ($1\,015$~yr, \citealt{Uchida2013Asymmetric1006,Winkler2014ARemnant}); and N103B = SNR 0509--68.7 (less than 850~yr, \citealt{Williams2018TheN103B}). We note that two of these SNRs, as observed in X-rays with Chandra, bear some similarity with our D$^6$ SNR: SNR 0519--69.0 looks mostly spherically symmetric except for a flatter section (similar to our model observed along the z-axis), N103B = SNR 0509-68.7 exhibits a darker circular region (similar to our model observed along the x-axis). 
Here we have to acknowledge the current limitations of both the simulations and the observations. Our simulations assume a clean environment, our maps are therefore an ideal case scenario. In the LMC and SMC, some SNRs have interesting features but are poorly resolved. We hope our work can help guide further careful analysis of nearby SNRs. We will dedicate a subsequent paper to spectroscopy diagnostics.

Finally, we note that one of the three hyper-velocity WDs found by \cite{Shen2018ThreeSupernovae} points back to a recently found SNR candidate G70.0--21.5, however this scenario implies an explosion about $100\,000$~yr ago. This SNR is due for a more thorough X-ray study of the ejecta, but identifying the morphological features presented in this work will probably not be possible for such an old object.

\section{Conclusion} 
\label{sec:conclusion}

In this paper, for the first time we have followed a ``helium-ignited violent merger'' or ``dynamically-driven double degenerate double detonation'' (D$^6$) SN model into the SNR phase, up to $4\,000$~yr after the explosion. We have analyzed the structure of the SNR using a variety of representations: 2D slices, 2D projections, 3D contours and their angular variations. We have found that a D$^6$ progenitor system and explosion leaves clear signatures on the SNR:
\begin{itemize}
\item the first detonation produces an ejecta tail, that at early times looks like a protrusion from the shell; 
\item the second detonation leaves a central density peak, which is revealed in X-rays when the RS reaches the centre;
\item because of the initial velocity shift, the SNR shell is off-centre at all times, which shows as a strong dipole component in the angular spectra;
\item the companion star generates a conical shadow in the ejecta, that is visible in projection as a dark patch surrounded by a bright ring.
\end{itemize}

Basically we found that the specific 3D structure of the explosion is preserved in the SNR phase. The features from the first detonation and from the companion are localized, and so the way they look depends on the direction of observation, producing various SNR morphologies. But since we see all the shocked material in projection in X-rays, they should be visible to some degree along any orientation. The features from the shadow are long lasting, they could in principle be detected in the shocked ejecta up to the time when the RS rebounds at the centre of the SNR, which happens just short of $2\,700$~yr after the explosion for an assumed density of 0.1 cm$^{-3}$. The conical shadow is visible on the SNR shell as a brighter ring encompassing a darker disk. The ring is a region of over-growth of the RTI, while the disk is a region of under-growth of it. This rather unusual SNR morphology is obtained even in a uniform ambient medium. So we point out that observing an irregular Type~Ia SNR does not necessarily implies an inhomogeneous medium.

We emphasize that we have only one realization of the $D^6$ model. 
The size of the shadow may depend on the strength of the initial interaction between the ejecta and the companion, which may depend on the masses and explosion energy. The relative orientation between the tail from the first detonation and the conical shadow from the companion may also vary. For instance, the tail may happen to be aligned with the shadow, which may affect its evolution.
Our work, which shows that the initial configuration matters in latter phases, should entice modellers to do more exhaustive studies of the $D^6$ scenario. 

In a follow-up paper, we will conduct a more advanced study of the thermodynamical state of the plasma and its thermal emission, in order to allow for more precise comparisons of our simulation results with spatially- and spectrally-resolved X-ray observations. Our additional perspectives with D$^6$ modelling include, for the early phase, to study the nebular emission by means of radiative transfer calculations, and for the later phase, to study the visibility of the RS at and after its rebound. The later point warrants further study regarding its observability in SNRs, not just for D$^6$. 
Finally we would like to encourage searches for WDs in SNRs: D$^6$ is an example of a thermonuclear explosion with a surviving companion that is not a normal star.


\begin{acknowledgments}
This research has been supported by Grants-in-Aid for Scientific Research (16K17656, 19K03907).
SN acknowledges the support by JSPS Grants-in-Aid for Scientific Research “KAKENHI” (A: Grant Number JP19H00693) and Pioneering Program of RIKEN for Evolution of Matter in the Universe (r-EMU).
This work was funded in part by the Interdisciplinary Theoretical and Mathematical Sciences (iTHEMS, \url{https://ithems.riken.jp}) program at RIKEN.
SSH acknowledges support by the Natural Sciences and Engineering Research Council of Canada (NSERC) and the Canadian Space Agency.
We thank the anonymous referee for comments that helped clarify the manuscript.
\\
\end{acknowledgments}

\vspace{5mm}

\facilities{
Simulation for the SN explosion was done on Oakforest-PACS at Joint Center for Advanced High Performance Computing and on Cray XC50 at Center for Computational Astrophysics, National Astronomical Observatory of Japan.
Simulations for the SNR evolution were performed on the iTHEMS clusters at RIKEN. 
}

\software{HEALPix \citep{Gorski2005HEALPixSphere}, SciPy \citep{Virtanen2020SciPyPython}, Matplotlib \citep{Hunter2007Matplotlib:Environment}, VisIt \citep{Childs2012VisIt:Data}}



\bibliography{references}{}

\begin{thebibliography}{}
\expandafter\ifx\csname natexlab\endcsname\relax\def\natexlab#1{#1}\fi
\providecommand{\url}[1]{\href{#1}{#1}}
\providecommand{\dodoi}[1]{doi:~\href{http://doi.org/#1}{\nolinkurl{#1}}}
\providecommand{\doeprint}[1]{\href{http://ascl.net/#1}{\nolinkurl{http://ascl.net/#1}}}
\providecommand{\doarXiv}[1]{\href{https://arxiv.org/abs/#1}{\nolinkurl{https://arxiv.org/abs/#1}}}

\bibitem[{{Benvenuto} {et~al.}(2015){Benvenuto}, {Panei}, {Nomoto}, {Kitamura},
  \& {Hachisu}}]{Benvenuto2015SpinUpSpinDown}
{Benvenuto}, O.~G., {Panei}, J.~A., {Nomoto}, K., {Kitamura}, H., \& {Hachisu},
  I. 2015, \apjl, 809, L6, \dodoi{10.1088/2041-8205/809/1/L6}

\bibitem[{Boehner {et~al.}(2017)Boehner, Plewa, \&
  Langer}]{Boehner2017ImprintsCompanions}
Boehner, P., Plewa, T., \& Langer, N. 2017, Monthly Notices of the Royal
  Astronomical Society, 465, 2060, \dodoi{10.1093/mnras/stw2737}

\bibitem[{Borkowski {et~al.}(2013)Borkowski, Reynolds, Hwang, Green, Petre,
  Krishnamurthy, \& Willett}]{Borkowski2013SupernovaG1.9+0.3}
Borkowski, K.~J., Reynolds, S.~P., Hwang, U., {et~al.} 2013, Astrophysical
  Journal Letters, 771, 2, \dodoi{10.1088/2041-8205/771/1/L9}

\bibitem[{Burkey {et~al.}(2013)Burkey, Reynolds, Borkowski, \&
  Blondin}]{Burkey2013X-RaySupernova}
Burkey, M.~T., Reynolds, S.~P., Borkowski, K.~J., \& Blondin, J.~M. 2013,
  Astrophysical Journal, 764, 63, \dodoi{10.1088/0004-637X/764/1/63}

\bibitem[{Cao {et~al.}(2015)Cao, Kulkarni, Howell, Gal-Yam, Kasliwal, Valenti,
  Johansson, Amanullah, Goobar, Sollerman, Taddia, Horesh, Sagiv, Cenko,
  Nugent, Arcavi, Surace, Wo{\'{z}}niak, Moody, Rebbapragada, Bue, \&
  Gehrels}]{Cao2015StrongSupernova}
Cao, Y., Kulkarni, S.~R., Howell, D.~A., {et~al.} 2015, Nature, 521, 328,
  \dodoi{10.1038/nature14440}

\bibitem[{Childs {et~al.}(2012)Childs, Brugger, Whitlock, Meredith, Ahern,
  Pugmire, Biagas, Miller, Harrison, Weber, Krishnan, Fogal, Sanderson, Garth,
  Bethel, Camp, R{\"{u}}bel, Durant, Favre, \&
  Navr{\'{a}}til}]{Childs2012VisIt:Data}
Childs, H., Brugger, E., Whitlock, B., {et~al.} 2012, {VisIt: An End-User Tool
  for Visualizing and Analyzing Very Large Data}

\bibitem[{Cioffi {et~al.}(1988)Cioffi, McKee, \&
  Bertschinger}]{Cioffi1988DynamicsRemnants}
Cioffi, D.~F., McKee, C.~F., \& Bertschinger, E. 1988, Astrophysical Journal,
  334, 252, \dodoi{10.1086/166834}

\bibitem[{Dessart {et~al.}(2020)Dessart, Leonard, \&
  Prieto}]{Dessart2020SpectralSupernova}
Dessart, L., Leonard, D.~C., \& Prieto, J.~L. 2020, A{\&}A, 638, A80,
  \dodoi{10.1051/0004-6361/202037854}

\bibitem[{{Di Stefano} {et~al.}(2011){Di Stefano}, {Voss}, \&
  {Claeys}}]{DiStefano2011SpinUpSpinDown}
{Di Stefano}, R., {Voss}, R., \& {Claeys}, J.~S.~W. 2011, \apjl, 738, L1,
  \dodoi{10.1088/2041-8205/738/1/L1}

\bibitem[{Dilday {et~al.}(2012)Dilday, Howell, Cenko, Silverman, Nugent,
  Sullivan, Ben-Ami, Bildsten, Bolte, Endl, Filippenko, Gnat, Horesh, Hsiao,
  Kasliwal, Kirkman, Maguire, Marcy, Moore, Pan, Parrent, Podsiadlowski,
  Quimby, Sternberg, Suzuki, Tytler, Xu, Bloom, Gal-Yam, Hook, Kulkarni, Law,
  Ofek, Polishook, \& Poznanski}]{Dilday2012PTF11kx:Progenitor}
Dilday, B., Howell, D.~A., Cenko, S.~B., {et~al.} 2012, Science, 337, 942,
  \dodoi{10.1126/science.1219164}

\bibitem[{Dwarkadas \& Chevalier(1998)}]{Dwarkadas1998InteractionSurroundings}
Dwarkadas, V., \& Chevalier, R.~A. 1998, Astrophysical Journal, 497, 807,
  \dodoi{10.1086/305478}

\bibitem[{{Eggleton}(1983)}]{Eggleton1983Aproximationslobes}
{Eggleton}, P.~P. 1983, \apj, 268, 368, \dodoi{10.1086/160960}

\bibitem[{Ferrand {et~al.}(2010)Ferrand, Decourchelle, Ballet, Teyssier, \&
  Fraschetti}]{Ferrand20103DAcceleration}
Ferrand, G., Decourchelle, A., Ballet, J., Teyssier, R., \& Fraschetti, F.
  2010, Astronomy {\&} Astrophysics, 509, L10,
  \dodoi{10.1051/0004-6361/200913666}

\bibitem[{Ferrand {et~al.}(2012)Ferrand, Decourchelle, \&
  Safi-Harb}]{Ferrand2012Three-dimensionalAcceleration}
Ferrand, G., Decourchelle, A., \& Safi-Harb, S. 2012, Astrophysical Journal,
  760, 34, \dodoi{10.1088/0004-637X/760/1/34}

\bibitem[{Ferrand {et~al.}(2014)Ferrand, Decourchelle, \&
  Safi-Harb}]{Ferrand2014Three-dimensionalAcceleration}
---. 2014, Astrophysical Journal, 789, 49, \dodoi{10.1088/0004-637X/789/1/49}

\bibitem[{Ferrand {et~al.}(2019)Ferrand, Warren, Ono, Nagataki, R{\"{o}}pke, \&
  Seitenzahl}]{Ferrand2019FromExplosion}
Ferrand, G., Warren, D.~C., Ono, M., {et~al.} 2019, The Astrophysical Journal,
  877, 136, \dodoi{10.3847/1538-4357/ab1a3d}

\bibitem[{Ferrand {et~al.}(2021)Ferrand, Warren, Ono, Nagataki, R{\"{o}}pke,
  Seitenzahl, Lach, Iwasaki, \& Sato}]{Ferrand2021FromModels}
---. 2021, The Astrophysical Journal, 906, 93, \dodoi{10.3847/1538-4357/abc951}

\bibitem[{Garc{\'{i}}a-Senz {et~al.}(2012)Garc{\'{i}}a-Senz, Badenes, \&
  Serichol}]{Garcia-Senz2012IsRemnants}
Garc{\'{i}}a-Senz, D., Badenes, C., \& Serichol, N. 2012, The Astrophysical
  Journal, 745, \dodoi{10.1088/0004-637X/745/1/75}

\bibitem[{Garc{\'{i}}a-Senz {et~al.}(2019)Garc{\'{i}}a-Senz, Velarde,
  Suzuki-Vidal, Stehl{\'{e}}, Cotelo, Portillo, Plewa, \&
  Pak}]{Garcia-Senz2019InteractionLaboratory}
Garc{\'{i}}a-Senz, D., Velarde, P., Suzuki-Vidal, F., {et~al.} 2019, The
  Astrophysical Journal, 871, 177, \dodoi{10.3847/1538-4357/aaf894}

\bibitem[{Garnavich(2017)}]{Garnavich2017DiscoveryAcceleration}
Garnavich, P. 2017, in Handbook of Supernovae (Cham: Springer International
  Publishing), 2605--2613, \dodoi{10.1007/978-3-319-21846-5{\_}104}

\bibitem[{Gorski {et~al.}(2005)Gorski, Hivon, Banday, Wandelt, Hansen,
  Reinecke, \& Bartelman}]{Gorski2005HEALPixSphere}
Gorski, K.~M., Hivon, E., Banday, A.~J., {et~al.} 2005, The Astrophysical
  Journal, 622, 759, \dodoi{10.1086/427976}

\bibitem[{Gray {et~al.}(2016)Gray, Raskin, \& Owen}]{Gray2016ShadowsRemnants}
Gray, W.~J., Raskin, C., \& Owen, J.~M. 2016, The Astrophysical Journal, 833,
  62, \dodoi{10.3847/1538-4357/833/1/62}

\bibitem[{Gronow {et~al.}(2020)Gronow, Collins, Ohlmann, Pakmor, Kromer,
  Seitenzahl, Sim, \& R{\"{o}}pke}]{Gronow2020SNeMechanism}
Gronow, S., Collins, C., Ohlmann, S.~T., {et~al.} 2020, Astronomy and
  Astrophysics, 635, 1, \dodoi{10.1051/0004-6361/201936494}

\bibitem[{{Gronow} {et~al.}(2021){Gronow}, {Collins}, {Sim}, \&
  {R{\"o}pke}}]{Gronow2021DoubleDetonations}
{Gronow}, S., {Collins}, C.~E., {Sim}, S.~A., \& {R{\"o}pke}, F.~K. 2021, \aap,
  649, A155, \dodoi{10.1051/0004-6361/202039954}

\bibitem[{{Guillochon} {et~al.}(2010){Guillochon}, {Dan}, {Ramirez-Ruiz}, \&
  {Rosswog}}]{Guillochon2010SurfaceDetonations}
{Guillochon}, J., {Dan}, M., {Ramirez-Ruiz}, E., \& {Rosswog}, S. 2010, \apjl,
  709, L64, \dodoi{10.1088/2041-8205/709/1/L64}

\bibitem[{{Hachisu} {et~al.}(2012){Hachisu}, {Kato}, \&
  {Nomoto}}]{Hachisu2012SpinUpSpinDown}
{Hachisu}, I., {Kato}, M., \& {Nomoto}, K. 2012, \apjl, 756, L4,
  \dodoi{10.1088/2041-8205/756/1/L4}

\bibitem[{Hillebrandt {et~al.}(2013)Hillebrandt, Kromer, R{\"{o}}pke, \&
  Ruiter}]{Hillebrandt2013TowardsObservations}
Hillebrandt, W., Kromer, M., R{\"{o}}pke, F.~K., \& Ruiter, A.~J. 2013,
  Frontiers of Physics, 8, 116, \dodoi{10.1007/s11467-013-0303-2}

\bibitem[{Hunter(2007)}]{Hunter2007Matplotlib:Environment}
Hunter, J.~D. 2007, Computing in Science {\&} Engineering, 9, 90,
  \dodoi{10.1109/MCSE.2007.55}

\bibitem[{Iben \& Tutukov(1984)}]{Iben1984Supernovaemass}
Iben, I.~J., \& Tutukov, A.~V. 1984, ApJS, 54, 335, \dodoi{10.1086/190932}

\bibitem[{{Jha} {et~al.}(2019){Jha}, {Maguire}, \&
  {Sullivan}}]{Jha2019Observationalsupernovae}
{Jha}, S.~W., {Maguire}, K., \& {Sullivan}, M. 2019, Nature Astronomy, 3, 706,
  \dodoi{10.1038/s41550-019-0858-0}

\bibitem[{{Ji} {et~al.}(2013){Ji}, {Fisher}, {Garc{\'\i}a-Berro}, {Tzeferacos},
  {Jordan}, {Lee}, {Lor{\'e}n-Aguilar}, {Cremer}, \&
  {Behrends}}]{Ji2013MagneticViscosity}
{Ji}, S., {Fisher}, R.~T., {Garc{\'\i}a-Berro}, E., {et~al.} 2013, \apj, 773,
  136, \dodoi{10.1088/0004-637X/773/2/136}

\bibitem[{Justham(2011)}]{Justham2011Single-degenerateContamination}
Justham, S. 2011, ApJL, 730, L34, \dodoi{10.1088/2041-8205/730/2/L34}

\bibitem[{Kasen(2010)}]{Kasen2010SeeingStar}
Kasen, D. 2010, The Astrophysical Journal, 708, 1025,
  \dodoi{10.1088/0004-637X/708/2/1025}

\bibitem[{Kashi \& Soker(2011)}]{Kashi2011Acircumbinarysupernovae}
Kashi, A., \& Soker, N. 2011, Mon. Not. R. Astron. Soc., 417, 1466,
  \dodoi{10.1111/j.1365-2966.2011.19361.x}

\bibitem[{{Kasuga} {et~al.}(2021){Kasuga}, {Vink}, {Katsuda}, {Uchida},
  {Bamba}, {Sato}, \& {Hughes}}]{Kasuga2021SpatiallyRemnant}
{Kasuga}, T., {Vink}, J., {Katsuda}, S., {et~al.} 2021, \apj, 915, 42,
  \dodoi{10.3847/1538-4357/abff4f}

\bibitem[{{Kelly} {et~al.}(2014){Kelly}, {Fox}, {Filippenko}, {Cenko}, {Prato},
  {Schaefer}, {Shen}, {Zheng}, {Graham}, \& {Tucker}}]{Kelly2014NearbySample}
{Kelly}, P.~L., {Fox}, O.~D., {Filippenko}, A.~V., {et~al.} 2014, \apj, 790, 3,
  \dodoi{10.1088/0004-637X/790/1/3}

\bibitem[{Kerzendorf {et~al.}(2018)Kerzendorf, Strampelli, Shen, Schwab,
  Pakmor, Do, Buchner, \& Rest}]{Kerzendorf2018A1006}
Kerzendorf, W.~E., Strampelli, G., Shen, K.~J., {et~al.} 2018, Monthly Notices
  of the Royal Astronomical Society, 479, 192, \dodoi{10.1093/mnras/sty1357}

\bibitem[{{Kopal}(1959)}]{Kopal1959Closebinarysystems}
{Kopal}, Z. 1959, {Close binary systems}

\bibitem[{Kosenko {et~al.}(2010)Kosenko, Helder, \& Vink}]{Kosenko2010TheStudy}
Kosenko, D., Helder, E.~A., \& Vink, J. 2010, Astronomy {\&} Astrophysics, 519,
  A11, \dodoi{10.1051/0004-6361/200913903}

\bibitem[{{Kromer} {et~al.}(2013){Kromer}, {Pakmor}, {Taubenberger}, {Pignata},
  {Fink}, {R{\"o}pke}, {Seitenzahl}, {Sim}, \&
  {Hillebrandt}}]{Kromer2013SN2010lp}
{Kromer}, M., {Pakmor}, R., {Taubenberger}, S., {et~al.} 2013, \apjl, 778, L18,
  \dodoi{10.1088/2041-8205/778/1/L18}

\bibitem[{{Kromer} {et~al.}(2016){Kromer}, {Fremling}, {Pakmor},
  {Taubenberger}, {Amanullah}, {Cenko}, {Fransson}, {Goobar}, {Leloudas},
  {Taddia}, {R{\"o}pke}, {Seitenzahl}, {Sim}, \&
  {Sollerman}}]{Kromer2016iPTF14atg}
{Kromer}, M., {Fremling}, C., {Pakmor}, R., {et~al.} 2016, \mnras, 459, 4428,
  \dodoi{10.1093/mnras/stw962}

\bibitem[{Li {et~al.}(2011)Li, Leaman, Chornock, Filippenko, Poznanski,
  Ganeshalingam, Wang, Modjaz, Jha, Foley, \& Smith}]{Li2011NearbySample}
Li, W., Leaman, J., Chornock, R., {et~al.} 2011, Monthly Notices of the Royal
  Astronomical Society, 412, 1441, \dodoi{10.1111/j.1365-2966.2011.18160.x}

\bibitem[{Liu \& Zeng(2021)}]{Liu2021Long-termSupernovae}
Liu, Z.-W., \& Zeng, Y. 2021, MNRAS, 500, 301, \dodoi{10.1093/MNRAS/STAA3280}

\bibitem[{Liu {et~al.}(2013)Liu, Pakmor, Seitenzahl, Hillebrandt, Kromer,
  R{\"{o}}pke, Edelmann, Taubenberger, Maeda, Wang, \&
  Han}]{Liu2013TheScenario}
Liu, Z.-W., Pakmor, R., Seitenzahl, I.~R., {et~al.} 2013, Astrophysical
  Journal, 774, \dodoi{10.1088/0004-637X/774/1/37}

\bibitem[{Lopez {et~al.}(2011)Lopez, Ramirez-Ruiz, Huppenkothen, Badenes, \&
  Pooley}]{Lopez2011UsingMixing}
Lopez, L.~A., Ramirez-Ruiz, E., Huppenkothen, D., Badenes, C., \& Pooley, D.~A.
  2011, Astrophysical Journal, 732, 114, \dodoi{10.1088/0004-637X/732/2/114}

\bibitem[{Lu {et~al.}(2011)Lu, Wang, Ge, Qu, Yang, Zheng, \&
  Chen}]{Lu2011TheCompanion}
Lu, F.~J., Wang, Q.~D., Ge, M.~Y., {et~al.} 2011, Astrophysical Journal, 732,
  11, \dodoi{10.1088/0004-637X/732/1/11}

\bibitem[{Maeda {et~al.}(2014)Maeda, Kutsuna, \&
  Shigeyama}]{Maeda2014SignaturesSupernovae}
Maeda, K., Kutsuna, M., \& Shigeyama, T. 2014, The Astrophysical Journal, 794,
  \dodoi{10.1088/0004-637X/794/1/37}

\bibitem[{Maoz {et~al.}(2014)Maoz, Mannucci, \&
  Nelemans}]{Maoz2014ObservationalSupernovae}
Maoz, D., Mannucci, F., \& Nelemans, G. 2014, Annual Review of Astronomy and
  Astrophysics, 52, 107, \dodoi{10.1146/annurev-astro-082812-141031}

\bibitem[{Marietta {et~al.}(2000)Marietta, Burrows, \&
  Fryxell}]{Marietta2000TypeIAConsequences}
Marietta, E., Burrows, A., \& Fryxell, B. 2000, Astrophysical Journal,
  Supplement, 128, 615, \dodoi{10.1086/313392}

\bibitem[{{Marion} {et~al.}(2016){Marion}, {Brown}, {Vink{\'o}}, {Silverman},
  {Sand}, {Challis}, {Kirshner}, {Wheeler}, {Berlind}, {Brown}, {Calkins},
  {Camacho}, {Dhungana}, {Foley}, {Friedman}, {Graham}, {Howell}, {Hsiao},
  {Irwin}, {Jha}, {Kehoe}, {Macri}, {Maeda}, {Mandel}, {McCully}, {Pandya},
  {Rines}, {Wilhelmy}, \& {Zheng}}]{Marion2016InteractionWithBinaryCompanion}
{Marion}, G.~H., {Brown}, P.~J., {Vink{\'o}}, J., {et~al.} 2016, \apj, 820, 92,
  \dodoi{10.3847/0004-637X/820/2/92}

\bibitem[{Moranchel-Basurto {et~al.}(2020)Moranchel-Basurto, Vel{\'{a}}zquez,
  de~Parga, Reynoso, Schneiter, \&
  Esquivel}]{Moranchel-Basurto2020SimulatedModel}
Moranchel-Basurto, A., Vel{\'{a}}zquez, P.~F., de~Parga, G.~A., {et~al.} 2020,
  Monthly Notices of the Royal Astronomical Society, 1538, 1531,
  \dodoi{10.1093/mnras/staa627}

\bibitem[{Nomoto(1982)}]{Nomoto1982AccretingSupernovae}
Nomoto, K. 1982, The Astrophysical Journal, 257, 780, \dodoi{10.1086/160031}

\bibitem[{{Nomoto} \& {Kondo}(1991)}]{Nomoto1991AccretionInducedCollapse}
{Nomoto}, K., \& {Kondo}, Y. 1991, \apjl, 367, L19, \dodoi{10.1086/185922}

\bibitem[{Pakmor {et~al.}(2011)Pakmor, Hachinger, R{\"{o}}pke, \&
  Hillebrandt}]{Pakmor2011ViolentSupernovae}
Pakmor, R., Hachinger, S., R{\"{o}}pke, F.~K., \& Hillebrandt, W. 2011,
  Astronomy and Astrophysics, 528, \dodoi{10.1051/0004-6361/201015653}

\bibitem[{Pakmor {et~al.}(2012)Pakmor, Kromer, Taubenberger, Sim, R{\"{o}}pke,
  \& Hillebrandt}]{Pakmor2012NormalBinaries}
Pakmor, R., Kromer, M., Taubenberger, S., {et~al.} 2012, Astrophysical Journal
  Letters, 747, L10, \dodoi{10.1088/2041-8205/747/1/L10}

\bibitem[{Pakmor {et~al.}(2013)Pakmor, Kromer, Taubenberger, \&
  Springel}]{Pakmor2013Helium-ignitedSupernovae}
Pakmor, R., Kromer, M., Taubenberger, S., \& Springel, V. 2013, Astrophysical
  Journal Letters, 770, L8, \dodoi{10.1088/2041-8205/770/1/L8}

\bibitem[{Pakmor {et~al.}(2008)Pakmor, R{\"{o}}pke, Weiss, \&
  Hillebrandt}]{Pakmor2008TheimpactCompanions}
Pakmor, R., R{\"{o}}pke, F.~K., Weiss, A., \& Hillebrandt, W. 2008, Astronomy
  {\&} Astrophysics, 489, 943, \dodoi{10.1051/0004-6361:200810456}

\bibitem[{{Pakmor} {et~al.}(2021){Pakmor}, {Zenati}, {Perets}, \&
  {Toonen}}]{Pakmor2021HybridHeCO}
{Pakmor}, R., {Zenati}, Y., {Perets}, H.~B., \& {Toonen}, S. 2021, \mnras, 503,
  4734, \dodoi{10.1093/mnras/stab686}

\bibitem[{Pan {et~al.}(2012)Pan, Ricker, \& Taam}]{Pan2012ImpactScenario}
Pan, K.-c., Ricker, P.~M., \& Taam, R.~E. 2012, The Astrophysical Journal, 151,
  \dodoi{10.1088/0004-637X/750/2/151}

\bibitem[{Papish {et~al.}(2015)Papish, Soker, Garc{\'{i}}a-Berro, \&
  Aznar-Sigu{\'{a}}n}]{Papish2015TheSupernova}
Papish, O., Soker, N., Garc{\'{i}}a-Berro, E., \& Aznar-Sigu{\'{a}}n, G. 2015,
  Monthly Notices of the Royal Astronomical Society, 449, 942,
  \dodoi{10.1093/mnras/stv337}

\bibitem[{{Petruk} {et~al.}(2021){Petruk}, {Kuzyo}, {Orlando}, {Pohl}, \&
  {Brose}}]{Petruk2021Magnetophase}
{Petruk}, O., {Kuzyo}, T., {Orlando}, S., {Pohl}, M., \& {Brose}, R. 2021,
  \mnras, \dodoi{10.1093/mnras/stab1319}

\bibitem[{{Polin} {et~al.}(2019){Polin}, {Nugent}, \&
  {Kasen}}]{Polin2019ObservationalPredictions}
{Polin}, A., {Nugent}, P., \& {Kasen}, D. 2019, \apj, 873, 84,
  \dodoi{10.3847/1538-4357/aafb6a}

\bibitem[{{Polin} {et~al.}(2021){Polin}, {Nugent}, \&
  {Kasen}}]{Polin2021NebularModels}
---. 2021, \apj, 906, 65, \dodoi{10.3847/1538-4357/abcccc}

\bibitem[{Rest {et~al.}(2005)Rest, Suntzeff, Olsen, Prieto, Smith, Welch,
  Becker, Bergmann, Clocchiatti, Cook, Garg, Huber, Miknaitis, Minniti,
  Nikolaev, \& Stubbs}]{Rest2005LightCloud}
Rest, A., Suntzeff, N.~B., Olsen, K., {et~al.} 2005, Nature, 438, 1132,
  \dodoi{10.1038/nature04365}

\bibitem[{Reynolds(2017)}]{Reynolds2017DynamicalRemnants}
Reynolds, S.~P. 2017, in Handbook of Supernovae (Cham: Springer International
  Publishing), 1981--2004, \dodoi{10.1007/978-3-319-21846-5{\_}89}

\bibitem[{Reynolds {et~al.}(2007)Reynolds, Borkowski, Hwang, Hughes, Badenes,
  Laming, \& Blondin}]{Reynolds2007AInteraction}
Reynolds, S.~P., Borkowski, K.~J., Hwang, U., {et~al.} 2007, Astrophysical
  Journal, Letters, 668, L135, \dodoi{10.1086/522830}

\bibitem[{Ruiter(2020)}]{Ruiter2020TypeOrigin}
Ruiter, A.~J. 2020, in Proceedings of IAU Symposium 357, Vol. 357, 1--15,
  \dodoi{10.1017/S1743921320000587}

\bibitem[{Ruiz-Lapuente(2019)}]{Ruiz-Lapuente2019SurvivingObservations}
Ruiz-Lapuente, P. 2019, New Astronomy Reviews, 85, id. 101523,
  \dodoi{10.1016/j.newar.2019.101523}

\bibitem[{{Saio} \& {Nomoto}(1985)}]{Saio1985OxygenNeonMagnesium}
{Saio}, H., \& {Nomoto}, K. 1985, \aap, 150, L21

\bibitem[{{Sato} {et~al.}(2015){Sato}, {Nakasato}, {Tanikawa}, {Nomoto},
  {Maeda}, \& {Hachisu}}]{Sato2015ViolentMerger}
{Sato}, Y., {Nakasato}, N., {Tanikawa}, A., {et~al.} 2015, \apj, 807, 105,
  \dodoi{10.1088/0004-637X/807/1/105}

\bibitem[{{Sato} {et~al.}(2016){Sato}, {Nakasato}, {Tanikawa}, {Nomoto},
  {Maeda}, \& {Hachisu}}]{Sato2016ViolentMerger}
---. 2016, \apj, 821, 67, \dodoi{10.3847/0004-637X/821/1/67}

\bibitem[{{Schulreich} \&
  {Breitschwerdt}(2022)}]{Schulreich2022RTIeROSITAbubbles}
{Schulreich}, M.~M., \& {Breitschwerdt}, D. 2022, \mnras, 509, 716,
  \dodoi{10.1093/mnras/stab2940}

\bibitem[{{Schwab} {et~al.}(2012){Schwab}, {Shen}, {Quataert}, {Dan}, \&
  {Rosswog}}]{Schwab2012MagneticViscosity}
{Schwab}, J., {Shen}, K.~J., {Quataert}, E., {Dan}, M., \& {Rosswog}, S. 2012,
  \mnras, 427, 190, \dodoi{10.1111/j.1365-2966.2012.21993.x}

\bibitem[{{Shen} {et~al.}(2021){Shen}, {Blondin}, {Kasen}, {Dessart},
  {Townsley}, {Boos}, \& {Hillier}}]{Shen2021RadiativeTransfer}
{Shen}, K.~J., {Blondin}, S., {Kasen}, D., {et~al.} 2021, \apjl, 909, L18,
  \dodoi{10.3847/2041-8213/abe69b}

\bibitem[{Shen {et~al.}(2018{\natexlab{a}})Shen, Kasen, Miles, \&
  Townsley}]{Shen2018Sub-Chandrasekhar-massRevisited}
Shen, K.~J., Kasen, D., Miles, B.~J., \& Townsley, D.~M. 2018{\natexlab{a}},
  The Astrophysical Journal, 854, 52, \dodoi{10.3847/1538-4357/aaa8de}

\bibitem[{Shen \& Schwab(2017)}]{Shen2017WaitDecays}
Shen, K.~J., \& Schwab, J. 2017, ApJ, 834, 180,
  \dodoi{10.3847/1538-4357/834/2/180}

\bibitem[{Shen {et~al.}(2018{\natexlab{b}})Shen, Boubert, G{\"{a}}nsicke, Jha,
  Andrews, Chomiuk, Foley, Fraser, Gromadzki, Guillochon, Kotze, Maguire,
  Siebert, Smith, Strader, Badenes, Kerzendorf, Koester, Kromer, Miles, Pakmor,
  Schwab, Toloza, Toonen, Townsley, \& Williams}]{Shen2018ThreeSupernovae}
Shen, K.~J., Boubert, D., G{\"{a}}nsicke, B.~T., {et~al.} 2018{\natexlab{b}},
  The Astrophysical Journal, 865, 15, \dodoi{10.3847/1538-4357/aad55b}

\bibitem[{Sim {et~al.}(2010)Sim, R{\"{o}}pke, Hillebrandt, Kromer, Pakmor,
  Fink, Ruiter, \& Seitenzahl}]{Sim2010DetonationsDwarfs}
Sim, S.~A., R{\"{o}}pke, F.~K., Hillebrandt, W., {et~al.} 2010, Astrophysical
  Journal Letters, 714, 52, \dodoi{10.1088/2041-8205/714/1/L52}

\bibitem[{{Tanikawa} {et~al.}(2015){Tanikawa}, {Nakasato}, {Sato}, {Nomoto},
  {Maeda}, \& {Hachisu}}]{Tanikawa2015ViolentMerger}
{Tanikawa}, A., {Nakasato}, N., {Sato}, Y., {et~al.} 2015, \apj, 807, 40,
  \dodoi{10.1088/0004-637X/807/1/40}

\bibitem[{Tanikawa {et~al.}(2018)Tanikawa, Nomoto, \&
  Nakasato}]{Tanikawa2018Three-DimensionalCompanion}
Tanikawa, A., Nomoto, K., \& Nakasato, N. 2018, The Astrophysical Journal, 868,
  90, \dodoi{10.3847/1538-4357/aae9ee}

\bibitem[{Tanikawa {et~al.}(2019)Tanikawa, Nomoto, Nakasato, \&
  Maeda}]{Tanikawa2019Double-DetonationMaterials}
Tanikawa, A., Nomoto, K., Nakasato, N., \& Maeda, K. 2019, The Astrophysical
  Journal, 885, 103, \dodoi{10.3847/1538-4357/ab46b6}

\bibitem[{{Taubenberger} {et~al.}(2013){Taubenberger}, {Kromer}, {Pakmor},
  {Pignata}, {Maeda}, {Hachinger}, {Leibundgut}, \&
  {Hillebrandt}}]{Taubenberger2013SN2010lp}
{Taubenberger}, S., {Kromer}, M., {Pakmor}, R., {et~al.} 2013, \apjl, 775, L43,
  \dodoi{10.1088/2041-8205/775/2/L43}

\bibitem[{Teyssier(2002)}]{Teyssier2002CosmologicalRAMSES}
Teyssier, R. 2002, Astronomy {\&} Astrophysics, 385, 337,
  \dodoi{10.1051/0004-6361:20011817}

\bibitem[{{Timmes} {et~al.}(2000){Timmes}, {Hoffman}, \&
  {Woosley}}]{Timmes2000Approx13}
{Timmes}, F.~X., {Hoffman}, R.~D., \& {Woosley}, S.~E. 2000, \apjs, 129, 377,
  \dodoi{10.1086/313407}

\bibitem[{{Timmes} \& {Swesty}(2000)}]{Timmes2000HelmholtzEoS}
{Timmes}, F.~X., \& {Swesty}, F.~D. 2000, \apjs, 126, 501,
  \dodoi{10.1086/313304}

\bibitem[{Uchida {et~al.}(2013)Uchida, Yamaguchi, \&
  Koyama}]{Uchida2013Asymmetric1006}
Uchida, H., Yamaguchi, H., \& Koyama, K. 2013, Astrophysical Journal, 771, 56,
  \dodoi{10.1088/0004-637X/771/1/56}

\bibitem[{Vigh {et~al.}(2011)Vigh, Vel{\'{a}}zquez, G{\'{o}}mez, Reynoso,
  Esquivel, \& Matias~Schneiter}]{Vigh2011AsymmetriesRemnants}
Vigh, C.~D., Vel{\'{a}}zquez, P.~F., G{\'{o}}mez, D.~O., {et~al.} 2011,
  Astrophysical Journal, 727, 32, \dodoi{10.1088/0004-637X/727/1/32}

\bibitem[{Vink(2017)}]{Vink2017X-RayRemnants}
Vink, J. 2017, in Handbook of Supernovae (Cham: Springer International
  Publishing), 2063--2086, \dodoi{10.1007/978-3-319-21846-5{\_}92}

\bibitem[{Virtanen {et~al.}(2020)Virtanen, Gommers, Oliphant, Haberland, Reddy,
  Cournapeau, Burovski, Peterson, Weckesser, Bright, van~der Walt, Brett,
  Wilson, Jarrod~Millman, Mayorov, Nelson, Jones, Kern, Larson, Carey, Polat,
  Feng, Moore, Vand~erPlas, Laxalde, Perktold, Cimrman, Henriksen, Quintero,
  Harris, Archibald, Ribeiro, Pedregosa, van Mulbregt, \&
  Contributors}]{Virtanen2020SciPyPython}
Virtanen, P., Gommers, R., Oliphant, T.~E., {et~al.} 2020, Nature Methods, 17,
  261, \dodoi{10.1038/s41592-019-0686-2}

\bibitem[{Warren \& Blondin(2013)}]{Warren2013Three-dimensionalSNRs}
Warren, D.~C., \& Blondin, J.~M. 2013, Monthly Notices of the Royal
  Astronomical Society, 429, 3099, \dodoi{10.1093/mnras/sts566}

\bibitem[{Warren \& Hughes(2004)}]{Warren2004Raising050967.5}
Warren, J.~S., \& Hughes, J.~P. 2004, The Astrophysical Journal, 608, 261,
  \dodoi{10.1086/392528}

\bibitem[{Warren {et~al.}(2005)Warren, Hughes, Badenes, Ghavamian, McKee,
  Moffett, Plucinsky, Rakowski, Reynoso, \&
  Slane}]{Warren2005Cosmic-RayObservations}
Warren, J.~S., Hughes, J.~P., Badenes, C., {et~al.} 2005, Astrophysical
  Journal, 634, 376, \dodoi{10.1086/496941}

\bibitem[{Webbink(1984)}]{Webbink1984Doublesupernovae}
Webbink, R.~F. 1984, The Astrophysical Journal, 277, 355,
  \dodoi{10.1086/161701}

\bibitem[{Whelan \& Iben(1973)}]{Whelan1973BinariesI}
Whelan, J., \& Iben, Icko, J. 1973, The Astrophysical Journal, 186, 1007,
  \dodoi{10.1086/152565}

\bibitem[{Williams {et~al.}(2018)Williams, Blair, Borkowski, Hendrick, Long,
  Petre, Raymond, Rest, Reynolds, Sankrit, Seitenzahl, \&
  Winkler}]{Williams2018TheN103B}
Williams, B.~J., Blair, W.~P., Borkowski, K.~J., {et~al.} 2018, The
  Astrophysical Journal Letters, 865, L13, \dodoi{10.1039/c1an15630f}

\bibitem[{Winkler {et~al.}(2014)Winkler, Williams, Reynolds, Petre, Long,
  Katsuda, \& Hwang}]{Winkler2014ARemnant}
Winkler, P.~F., Williams, B.~J., Reynolds, S.~P., {et~al.} 2014, Astrophysical
  Journal, 781, 65, \dodoi{10.1088/0004-637X/781/2/65}

\bibitem[{Yamaguchi {et~al.}(2017)Yamaguchi, Hughes, Badenes, Bravo,
  Seitenzahl, Mart{\'{i}}nez-Rodr{\'{i}}guez, Park, \&
  Petre}]{Yamaguchi2017TheRemnant}
Yamaguchi, H., Hughes, J.~P., Badenes, C., {et~al.} 2017, The Astrophysical
  Journal, 834, 1, \dodoi{10.3847/1538-4357/834/2/124}

\bibitem[{Zeng {et~al.}(2020)Zeng, Liu, \& Han}]{Zeng2020TheStar}
Zeng, Y., Liu, Z.-W., \& Han, Z. 2020, ApJ, 898, 12,
  \dodoi{10.3847/1538-4357/ab9943}

\end{thebibliography}
\bibliographystyle{aasjournal}

\end{document}